\documentclass[oldversion]{aa}
\usepackage[T1]{fontenc}
\usepackage{psfig}
\usepackage{subfigure}
\usepackage{graphicx}
\usepackage{latexsym}
\usepackage{mathrsfs}
\usepackage{amsmath}
\usepackage{amsfonts}
\usepackage{natbib}
\usepackage{ulem}
\bibpunct{(}{)}{;}{a}{}{,}

\advance \voffset by 1.00cm\relax
\def\msun{{\,M_{\odot}}}

\def\mdot{\dot M}

\newcommand{\der}[2]{\ensuremath{\frac{{\rm d} #1}{{\rm d} #2}}}

\newcommand{\pder}[2]{\ensuremath{\frac{\partial #1}{\partial #2}}}

\newcommand{\derln}[2]{\ensuremath{\frac{{\rm d\,ln}\, #1}{{\rm d\,ln}\, #2}}}
\newcommand{\pderln}[2]{\ensuremath{\frac{\partial\,\rm ln\,#1}{\partial\,\rm ln\,#2}}}

\newcommand{\be}{\begin{equation}}
\newcommand{\ee}{\end{equation}}
\newcommand{\bea}{\begin{eqnarray}}
\newcommand{\eea}{\end{eqnarray}}



\begin{document}

\title{Relativistic slim disks with vertical structure}
    \author{
           Aleksander S\k{a}dowski \inst{1}
           \and
               Marek Abramowicz\inst{1,2}
           \and
               Michal Bursa \inst{3}
           \and
              W{\l}odek Klu{\'z}niak\inst{1}
           \and
               \\Jean-Pierre Lasota\inst{4,5}
           \and
               Agata R\'o{\.z}a{\'n}ska\inst{1}
           }
   \institute{
 Nicolaus Copernicus Astronomical Center, Polish Academy
             of Sciences,
             Bartycka 18, PL-00-716 Warszawa, Poland \\
             \email{as@camk.edu.pl}, ~\email{wlodek@camk.edu.pl}, ~\email{agata@camk.edu.pl}
         \and
             Department of Physics, G\"oteborg University,
             SE-412-96 G\"oteborg, Sweden    \\
             \email{Marek.Abramowicz@physics.gu.se}              
         \and
             Astronomical Institute, Academy of Sciences of the
            Czech Republic,
            Bo{\v c}ni II/1401a, 141-31  Prague,
            Czech Republic\\
             \email{bursa@astro.cas.cz}         
         \and
             Institut d'Astrophysique de Paris, UMR 7095 CNRS, UPMC Univ Paris 06, 98bis Bd Arago, 75014 Paris, France\\
             \email{lasota@iap.fr}
          \and
             Jagiellonian University Observatory, ul. Orla 171,
             PL-30-244 Krak{\'o}w,
             Poland          
            }
\abstract{We report on a scheme for incorporating vertical radiative energy
transport into a fully relativistic, Kerr-metric model of optically thick,
advective, transonic alpha disks.
Our code couples the radial and vertical equations of the accretion disk.
The flux was computed in the diffusion approximation,
and convection is included in the mixing-length approximation.
We present the detailed structure of this
``two-dimensional'' slim-disk model for $\alpha=0.01$.
We then calculated the emergent spectra
integrated over the disk surface. 
The values of surface density, radial velocity,
 and the photospheric height for these models differ by 20\%-30\% 
from those obtained in the polytropic, 
height-averaged slim disk model considered previously.
However, the emission profiles and the resulting
spectra are quite similar for both types of models.
The effective optical depth of the slim disk becomes lower than unity 
for high values of the alpha parameter and for high accretion rates.
}
\authorrunning{A. S{\k a}dowski et al.}
\titlerunning{Vertical structure of slim disks}
\maketitle

\section{Introduction}

Modeling accretion flows onto black holes is crucial for understanding
the energetic emissions observed from many sources, both Galactic and
extragalactic (e.g., X-ray binaries, ULXs, AGNs).  The accreting matter commonly
settles into a disk-like configuration in which angular momentum is
transported outwards as a result of shear in the differentially
rotating fluid, allowing mass to be accreted onto the central compact object. 
Of particular interest are sources such as the
low-mass X-ray binaries in which matter is transferred onto the X-ray source
from a binary companion, usually a late-type star. 
When containing a black hole, these sources
undergo irregular outbursts (lasting several months or years) that are
analogous to the outbursts observed in dwarf novae, and are thought to
be related to the viscous-thermal instability, which leads to an
enhanced transfer of angular momentum in the accreting fluid
\citep{lasota01}.
 
In outburst, the source is a luminous X-ray emitter and as the
luminosity decays from the outburst maximum to minimum, the disk is
thought to follow a sequence of quasi-stationary states, whose spectra
can be derived from steady-state models of accretion disks.  At high
luminosities describing the inner regions of such accretion
disks by the most-often used ``thin disk" models is no longer valid. A
description of such disks in terms of  ``slim disk" models
 is more appropriate.

Following the approach of \cite{shakura-73},
in the ``standard" discussion of stationary disk structure one assumes 
that there is no radial
advection of heat (the flow is ``radiatively efficient''), the radial
pressure gradients are negligible, and the distribution of angular
momentum is Keplerian. All of these assumptions hold for very
thin disks, including the original \cite{shakura-73} model (henceforth SS)
and its general-relativistic version (NT) which was developed for 
the Kerr geometry
by \cite{nt}.  However, it has long been realized that not all disk-like
accretion flows satisfy these assumptions  In particular, slim disks
\citep[e.g.,][]{slim} form a stable branch of accretion, in
which advection of entropy is important and the disk is not necessarily
geometrically thin. These solutions are particularly relevant to
sources with high accretion rates, and they converge to the standard case
in the low accretion rate limit.

In the case of thin disks, methods for treating the radiative transfer
and computing the vertical structure of the disk
 have been developed by \cite{shawehrse86} and \cite{hubeny91}
and applied by, e.g.,
\cite{davishubeny-06}, \cite{idanlasota08}, and \cite{rozanskamadej08}.
These have not yet been applied to slim disks. Until recently,
the properties of slim disk models have been represented with
quantities averaged over the vertical structure of the disk.
 This paper presents improved advective
steady-state models of accretion, in which the vertical structure of
the ``slim disk'' is explicitly taken into account.

Recently, \cite{sadowski.slim} revisited slim disk models with an
improved numerical code, while \cite{sadowski.photosphere} performed a
preliminary study of the slim disk vertical structure. While our
present work is based on these two papers, it breaks with the long
tradition of computing the radial dependence of disk quantities,
including the emergent flux $F(r)$, before the vertical structure is analyzed.
In this paper, by closely coupling the radial
properties of the disk to its vertical structure, we offer a more
consistent treatment of the slim disk.
A similar approach has been taken by \cite{nir} in their discussion of winds
from Super-Eddington slim disks in a pseudo-Newtonian potential.

The high luminosities observed in the bright X-ray sources imply there 
is an effective mechanism of angular momentum transport in
the accreting fluid, and much work has been devoted to MHD simulations
of one such mechanism, the magneto-rotational instability (MRI). For
thin disks, the effectiveness of angular momentum transport is
conveniently parametrized by the alpha parameter that was introduced
by Shakura and Sunyaev (1973).  Observations of systems in which
variability is driven by accretion disk instabilities suggest rather
high values of $\alpha$ for fully ionized disks, much higher than the
value  $\alpha=0.001$ that was adopted in the original slim-disk
article \citep{slim}. \citet{smak99} has already showed that one has to take $\alpha\approx 0.2$ to describe
hot disks in dwarf novae. As
reviewed by \citet{kingetal07}, the value of $\alpha$ in hot, fully
ionized disks has to be $\sim 0.1 - 0.4$, whereas in cold protostellar
and FU Urionis disks, much lower values of this parameter are required
$\sim 0.01$ and $\sim 0.001 - 0.003$, respectively. Numerical
simulations give a value of $\alpha \sim 0.01$ for ionized disks, ten
times lower than suggested by observations, and it
probably reflects the limitations of the present MRI calculations
\citep{kingetal07}.  However, the apparently observational
determinations of $\alpha$ mentioned above are in fact strongly
model-dependent. In the case of dwarf-novae, for example, these
determinations assume that outbursts are described well by the
thermal-viscous instability model. However, the case of the epitome
dwarf-nova SS Cyg shows that such a description might be inadequate
\citep{schreiberlasota-07, smak-10}, putting presumably
observational determinations into doubt.  Therefore, one cannot exclude that
the numerical simulations are correct after all.

In this paper we adopt the value $\alpha=0.01$. With such a low value
of $\alpha$,  the resulting
disk models are optically thick ($\tau_{\rm eff}>1$), thus allowing a
simplified treatment of radiative transfer. We treat the vertical
energy transport in the diffusive approximation for the radiative flux
(and account for vertical convection in the mixing-length
approximation).

We begin the body of the paper with a brief discussion of the model,
then both the radial and the vertical equations of disk structure are
presented. In Sect.~\ref{s.numerical} we describe the  numerical
method used to solve the problem. The radial structure of the
solutions is described in Sect.~\ref{ss.radialstructure}, while the
vertical profiles of various disk quantities are discussed in
Sect.~\ref{ss.vertstructure}. By comparing our results with previous
work, we find that, with a judicious choice of parameters, a standard
polytropic slim disk model can be made to approximate our full
solution. Analytical formulae for the appropriate values of the
parameters can be found in  Sect.~\ref{s.comparison}.  Finally, in the
last section we discuss the limitations of the model and its possible
applications.

\section{Slim-disk structure}
\label{s.structure}

Slim disks are alpha disks. Like Shakura-Sunyaev disks
they are based on the assumption that the dissipation mechanisms
operating in accretion flows may be described by a viscous stress
tensor whose leading component is proportional to the pressure.

Slim disk solutions, again just like the SS and NT
solutions, are obtained by solving vertically averaged (or
height-integrated), radial equations of motion. Thus, steady slim
disks, like the SS and NT ones, neglect the vertical structure of
flow, and describe essentially flat fluid configurations. Although an
expression for the radial dependence of disk thickness is obtained,
the slope of the disk surface is usually neglected in the discussion
of emergent spectra, where a plane-parallel atmosphere is typically
considered for the purposes of computing radiative transfer in the
vertical direction.

In the case of slim disks, the radial equations of structure
are a set of ordinary differential equations (ODEs).
For a detailed account of these Kerr-metric equations
(and state-of-the-art traditional slim-disk models),
see \cite{sadowski.slim}, who
followed previous work beginning with \cite{lasota94}, and considered 
subsequent improvements \citep{adafs, vertical, gammie}.
The set of equations is closed by including relations describing
vertical hydrostatic equilibrium and vertical transport of energy. By
solving these equations with appropriate boundary (or regularity)
conditions, one obtains the radial profiles of the central temperature,
$T_c(r)$, surface density, $\Sigma(r)$, the height-integrated
pressure, $P(r)$, the half-thickness of the disk, $h(r)$, a radial
velocity $V(r)$, the emerging flux of radiation, $F(r)$, and certain
other physical quantities. 

In this paper we are taking a first step towards constructing truly
three-dimensional slim disk models, by solving a set of differential
equations describing the vertical structure of the disk. The resulting
$z$-dependence of physical quantities is used to compute certain
coefficients that enter the radial equations and that up till now
have been estimated with algebraic expressions. 
We refer to the resulting models as
``two-dimensional (2-D) slim disk'' solutions,
although it has to be understood that in contrast to the thin-disk solutions of 
\cite{urpin} and \cite{kita}, the actual meridional flow has not
been computed in this paper, but the average radial velocity
alone has been considered in the structure equations.
However, the models are now
self-consistent in the sense that the vertical averages of physical
quantities that form the coefficients of the radial ODEs do correspond
to the vertical structure considered in the radiative transfer
calculation---in previous work the vertical structure of the radiative
atmosphere was considered {\it a posteriori}, and it had no influence
on the radial structure of the disk.

\subsection{Basic assumptions, parameters, and coefficients}
\label{s.assump}

We assume an axially symmetric, stationary fluid configuration in the
Kerr metric, with fixed values of the black hole (BH) mass, $M$, 
and spin, $a$, parameters and the
fluid disk is symmetric  under reflection in the equatorial plane of
the metric. Matter is supplied at a steady rate, $\dot M$, through a
boundary ``at infinity'' and angular momentum is removed through the
same boundary (in practice, we use the NT solution for the outer boundary
condition), whereas zero torque is assumed at the BH
horizon. We assume that no mass or angular momentum
crosses the disk surface. We neglect the loss of angular momentum to both wind and radiation. We do not consider self-irradiation of the disk
and assume that the magnetic pressure may be neglected. 
Neglecting the incoming radiation may not be justified for super-Eddington accretion rates
for which the disk is geometrically thick. 

A fraction, $(1+f^{\rm adv}(r))^{-1}$, of the entropy
generated locally by dissipative processes is released into the
radiation field, while the remainder is advected by the gas.

A unique solution to the slim-disk model can only be found if certain
additional assumptions are made. We make the following arbitrary
choice.  We neglect the vertical variation ($z$ dependence) of the
velocity field, considering only its height-averaged value; thus, the
velocity is always directed radially inwards and is a function of the
radial coordinate alone. Similarly, we assume there is no $z$
variation of the advection factor $f^{\rm adv}(r)$.  Dissipation and
angular momentum transport are given by the alpha prescription
\citep{shakura-73}, with a constant value of $\alpha$. We assume that
the dissipation rate is proportional to the total pressure,
$p$. For a more detailed statement, see Eq.~(\ref{vs.dFdz})
and the comment following it.
Calculations are carried out for the value $\alpha=0.01$.

We are looking for 2-D
slim-disk solutions at a definite value of mass
accretion rate for a given Kerr metric. Thus, for a fixed value of
$\alpha$, there are three fundamental parameters describing a given
slim-disk solution: $M$, $a$, and $\dot M$.

In the structure equations,  we take $G=c=1$ and make use of the following expressions involving the BH spin:
\begin{eqnarray}
\Delta&=&r^2-2Mr+a^2,\\\nonumber
A&=&r^4+r^2a^2+2Mra^2,\\\nonumber
\cal C&=&1-3r_*^{-1}+2a_*r_*^{-3/2}\\
\cal D&=&1-2r_*^{-1}+2a_*^2r_*^{-2}\\\nonumber
\cal H&=&1-4a_*r_*^{-3/2}+3a_*^{2}r_*^{-2} \nonumber
\end{eqnarray}
with $a_*=a/M$ and $r_*=r/M$.

We use the Boyer-Lindquist system of coordinates and introduce
  the vertical coordinate $z=r \cos\theta$. The metric near the
  equatorial plane takes the form \citep{adafs},
\be
ds^2=-\frac{r^2\Delta}{A}dt^2+\frac{A}{r^2}(d\phi - \omega dt)^2 +
\frac{r^2}\Delta dr^2+dz^2,
\ee
where $\omega=2Mar/A$ is the angular velocity of the frame dragging.

The radial gas (three-)velocity, $V$, as measured by an observer co-rotating
with the fluid at a fixed value of $r$, is given by the relation \citep{adafs}
\be
V/\sqrt{1-V^2}=u^r g_{rr}^{1/2}=\frac{r u^r}{\Delta^{1/2}},
\label{masscon}
\ee
where $u^r$ is the contravariant radial component of the 
fluid 4-velocity and has the dimension of physical velocity.
The disk surface density is $\Sigma=\int_{-h}^{+h}\rho\,dz$,
while the height-integrated pressure is $P=\int_{-h}^{+h}p\,dz$.
The total pressure is the sum of gas and radiation pressures, 
$p=p_{\rm gas}+p_{\rm rad}$. 
We adopt an equation of state corresponding to the choice
$p_{\rm gas}=k\rho T/(\mu m_{\rm p})$, and $p_{\rm rad}= aT^4/3$,
with $k$ the Boltzmann constant, $m_{\rm p}$ the proton mass,
and $a$ the radiation constant
(no confusion with the spin parameter may arise). The mean molecular weight is
taken to be $\mu=0.62$, but see the comment following Eq.~(\ref{vs.eqy2}).
The height-integrated energy density is
\be
E=\int_{-h}^{h}\left(\frac{p_{\rm gas}}{\gamma-1}+3p_{\rm rad}\right)\,dz,
\ee
where $\gamma=5/3$.

The following averages (moments)
enter the radial equations as coefficients of certain terms:
\begin{eqnarray}
\label{e.h1}
\eta_1&\equiv&\frac1{T_c^4}\int_{0}^{+h}T^4\,dz,\\
\label{e.h2}
\eta_2&\equiv&\frac2{\Sigma \,T_c}\int_{0}^{+h}\rho T\,dz,\\
\eta_3 &\equiv& E/P,\\
\eta_4&\equiv&\frac1\Sigma\int_{0}^{h}\rho z^2\,dz.
\end{eqnarray}
All integrals in this section are taken at a fixed value of $r$. 
Here, $T(z)$ is the gas temperature,
and $T_c$ is its value at the equatorial plane, $T(0)=T_c$.


The vertical epicyclic frequency squared, which can be thought of as
the vertical component of gravity, is \citep{kato-93},
\be
\Omega_\perp^2\equiv\frac{M}{r^3}\frac{\cal H}{\cal C}.
\label{eq.omegatilde}
\ee
We also define 
 \be
 \tilde\Gamma_1=1+\frac1{\eta_3}.
 \ee

\subsection{Vertical structure equations}
\label{s.verticaleq}
We describe the vertical structure of an accretion disk in the
optically thick regime by the following equations:

(i) Hydrostatic equilibrium \citep{KatoBook},
\begin{equation}
 \frac 1{\rho}\der pz=-\Omega_\perp^2z,
\label{vs.dpdz}
\end{equation}
where $\Omega_\perp$ is defined in Eq.~(\ref{eq.omegatilde}).
Although other expressions for the right-hand side of
Eq.~(\ref{vs.dpdz}) can be found in the literature (e.g., \cite{vertical}
includes $v^z\neq 0$), the form above is appropriate for our scheme, in which the vertical structure is precalculated before any information about the radial variables becomes available. Thus,
in Eqs. (\ref{eq.omegatilde}) and (\ref{vs.dpdz}), we assume Keplerian angular velocity $[M/({\cal C} r^3)]^{1/2}$ , as well as $v_z=0$ in hydrostatic equilibrium.

(ii) The energy generation equation. 
We assume that the vertical flux of energy inside the disk
$\cal F$ is generated
according to
\begin{equation}
\label{vs.dFdz}
 \der {\cal F}z=\frac{3\cal D}{2\cal C}\cdot
\left(\frac{\alpha p}{1+f^{\rm adv}}\right)\left(\frac{M}{r^3}\right)^{1/2}.
\end{equation}
Strictly speaking, this does not correspond to a constant $\alpha$ prescription,
as the term $(3/2)(M/r^3)^{1/2}$, which is derived from Keplerian strain,
departs somewhat from the value that would follow from the actually computed
distribution of angular momentum (cf., Fig.~\ref{f.angmom.a0}). However,
as the departure for sub-Eddingtonian accretion rates is small,
we expect Eq.~(\ref{vs.dFdz}) to afford a good approximation.
 Note that
$f^{\rm adv}=0$ corresponds to the NT disk
(going over into the Shakura-Sunyaev disk
in the nonrelativistic limit of thin disks),
$f^{\rm adv}>1$
characterizes advection-dominated disks, while $f^{\rm adv}<0$ describes
those disk regions where the advected heat is being released. The amount
of heat advected $Q^{\rm adv}=f^{\rm adv}{\cal F}(h)$.

(iii) Energy transport. The structure of the disk
has to be such that the actual value of the divergence
of the flux corresponds to Eq.~(\ref{vs.dFdz}).
Radiative transport is computed in the diffusive approximation
\begin{equation}
 {\cal F}(z)=-\frac{16\sigma T^3}{3\kappa\rho}\der Tz,
\end{equation}
while convective transport is computed in a mixing-length
approximation.
Energy is transported in the vertical direction
through diffusion of radiation or convection according
to the value of the thermodynamical gradient,
which can be either radiative or convective.
 Accordingly, we take
\begin{equation}
\derln Tp=\left\{\begin{array}{lll}
\nabla_{\rm rad}, & {\rm for} & \nabla_{\rm rad}\le\nabla_{\rm ad} \\
\nabla _{\rm conv}, & {\rm for} & \nabla_{\rm rad}>\nabla_{\rm ad} \\
\end{array}\right.
\label{vs.gradient}
\end{equation}
with the adiabatic gradient given by a derivative at constant entropy:
$\nabla_{\rm ad} \equiv (\partial\, {\rm ln}\,T/\partial\, {\rm ln}\,p)_S$.

The radiative gradient $\nabla_{\rm rad}$ is calculated in the
diffusive approximation,
\begin{equation}
\nabla_{\rm rad}=\frac{3p\kappa_R {\cal F}}{16\sigma T^4\Omega_\perp^2z},
\label{vs.gradrad}
\end{equation}
where $\kappa_R$ is the Rosseland mean opacity, and
$\sigma$ is the Stefan-Boltzmann constant. At the equatorial plane we
apply the boundary conditions described in point (iv) below.

When the temperature gradient exceeds the
value of the adiabatic gradient, we have to consider
the convective energy flux. The convective gradient
$\nabla _{\rm conv}$ is calculated using the mixing length theory
introduced by \citet{paczynski69}. We take the following
mixing length,
\begin{equation}
H_{\rm ml}=1.0 H_p,
\label{vs.hml}
\end{equation}
with pressure scale height $H_p$ defined as \citep{hameury98}
\begin{equation}
H_{p}=\frac p{\rho\Omega_\perp^2 z+\sqrt{p\rho}\Omega_\perp}
\label{vs.hP}.
\end{equation}
The convective gradient is defined by the formula
\begin{equation}
\nabla _{\rm conv}=\nabla_{\rm ad}+(\nabla_{\rm rad}-\nabla_{\rm ad})y(y+w)
\label{vs.gradconv}
\end{equation}
where $y$ is the solution of the equation
\begin{equation}
\frac94\frac{\tau_{ml}^2}{3+\tau_{ml}^2}y^3+wy^2+w^2y-w=0,
\label{vs.eqy1}
\end{equation}
with the typical optical depth for convection
$\tau_{ml}=\rho\kappa_R H_{ml}$, and $w$ given by

\begin{scriptsize}
\begin{equation}
\frac{1}{w^2}=\left(\frac{3+\tau_{ml}^2}{3\tau_{ml}}\right)^2
\frac{\Omega_\perp^2zH_{ml}^2\rho^2C_p^2}{512\sigma^2T^6H_P}
\left(\pderln\rho T\right)_p
\left(\nabla_{\rm rad}-\nabla_{\rm ad}\right)
\label{vs.eqy2}.
\end{equation}\end{scriptsize}
The thermodynamical quantities $C_p$, $\nabla_{\rm ad}$ and $(\partial\,{\rm ln}\,\rho/\partial\,{\rm ln}\,T)_p$ are calculated using
standard formulae \citep[e.g.,][]{stellarstructure} assuming solar abundances
($X=0.70$, $Y=0.28$) and, when necessary, taking the
effect of partial ionization of gas on the gas mean molecular
weight into account .

We use Rosseland mean opacities $\kappa_R$ (including 
the processes of absorption and scattering) taken from
\citet{alexander} and \citet{seaton}. Following other authors
\citep[e.g.,][]{idanlasota08}, we neglect expansion opacities,
in agreement with our neglect of vertical velocity gradients.

(iv) We set the following boundary conditions. At the equatorial plane
($z=0$) we set ${\cal F}(0)=0$, in accordance with the assumption of
reflection symmetry, while at the disk surface, $\tau(h)=0$, we follow
the Eddington approximation \citep{MihalasBook} and require
${\cal F}(h)\equiv\sigma T_{\rm eff}^4=2\sigma T^4(h)$.

In practice, for a fixed $r$, and prescribed values of $T_c$ and
$f^{\rm adv}$, a trial value of the central density, $\rho_c$, is assumed
and the equations are integrated in $z$ until
$\rho(z_*)=10^{-16}\,{\rm g\,cm^{-3}}$ (as a stand-in for the disk
surface, $z_*=h$). If ${\cal F}(h)$ and $T(h)$ fail to satisfy the surface
boundary condition, the assumed value of $\rho_c$ is adjusted, and the integration
is repeated, until the condition ${\cal F}(h)=2\sigma T^4(h)$ is
met. Convergence is usually attained in a few iterations.
The emergent flux of radiation at any given $r$ is then
$F={\cal F}(h)$.

\subsection{Radial structure equations}
\label{s.radialeq}

The radial sector of the model is described by four laws of
conservation and a regularity condition:

(i) Mass conservation,
\begin{equation}
 \dot M=-2\pi \Sigma r u^r.
\label{eq_cont2}
\end{equation}

(ii) Conservation of angular momentum,
\begin{equation}
 \frac{\dot{M}}{2\pi}({\cal L}-{\cal L}_{\rm in})=\frac{A^{1/2}\Delta^{1/2}\Gamma}{r}\alpha P,
\label{eq_ang6}
\end{equation}
where ${\cal L}=u_\phi$ is the specific angular momentum, ${\cal L}_{\rm in}$ is a constant, whose value is to be specified later, and 
$\Gamma$ is the Lorentz factor \citep{gammie}:
$$\Gamma^2=\frac1{1-V^2}+\frac{{\cal L}^2r^2}{A}.$$

(iii) Conservation of radial momentum,
\begin{equation}
\frac{V}{1-V^2}\frac{dV}{dr}=\frac{\cal A}{r}-\frac{1}{\Sigma}\frac{dP}{dr},
\label{eq_rad3}
\end{equation}
where
\begin{equation}
{\cal A}=-\frac{MA}{r^3\Delta\Omega_k^+\Omega_k^-}\frac{(\Omega-\Omega_k^+)(\Omega-\Omega_k^-)}{1-\tilde\Omega^2\tilde R^2},
\label{eq_rad4}
\end{equation}
and $\Omega=u^\phi /u^t$ is the angular velocity with respect to a stationary observer, $\tilde\Omega=\Omega-\omega$ is the angular velocity with respect to an inertial observer, $\Omega_k^\pm=\pm M^{1/2}/(r^{3/2}\pm aM^{1/2})$ are the angular frequencies of the corotating and counterrotating Keplerian orbits and $\tilde R=A/(r^2\Delta^{1/2})$ is the radius of gyration.

The value of $P$ is taken from vertical structure solutions for given values of $\Sigma$ and $T_c$, 
\begin{equation}
\label{e.P}
P=\eta_2\frac{k}{\mu m_{\rm p}}\Sigma T_c+\eta_1\frac23aT_c^4 ;
\end{equation}
hence, the radial derivative of the height-integrated pressure takes the form
\begin{eqnarray}
\nonumber
\label{e.dPdr}
\der{\ln P}r&=&(4-3\beta)\der{\ln T_c}r+\beta\der{\ln\Sigma}r+\\
&+&(1-\beta)\der{\ln \eta_1}r+\beta\der{\ln \eta_2}r,
\end{eqnarray}
with $\beta=\eta_2 (k/\mu m_{\rm p})\Sigma T_c/P$.

(iv) Energy conservation.\\
The advective cooling is defined following \cite{KatoBook} in terms of the vertically integrated quantities, as
\be
Q^{\rm adv}=\frac1r\der{}{r}(r u^r(E+P))-u^r\der Pr-\int_{-h}^{h}u^z\pder pz\,dz.
\label{eq.advcooling}
\ee

Using mass conservation (Eq. [\ref{eq_cont2}]) and hydrostatic equilibrium
(Eq. [\ref{vs.dpdz}]), the expression can be rewritten as
\begin{eqnarray}\nonumber
\label{eq.qadv}
Q^{\rm adv}&=&-\frac{\mdot}{2\pi r^2}\left(\eta_3\frac{P}\Sigma\derln{P}{r} - (1+\eta_3)\frac{P}\Sigma\derln\Sigma r+\right.\\
&+&\left. \eta_3\frac P\Sigma\derln{\eta_3}{r}+\Omega^2_\perp\eta_4\derln{\eta_4}{r}\right).
\end{eqnarray}
Just like $P$, the advective cooling term, $Q^{\rm adv}$, and the coefficients
 $\eta_3$, $\eta_4$ are all determined from the vertical structure solutions.

The manipulation of the last term in Eq.~(\ref{eq.advcooling}) was as follows:
\begin{eqnarray}\nonumber
&&-\int_{-h}^{h}u^z\pder pz\,dz=\Omega^2_\perp\int_{-h}^{h}u^z\rho z\,dz=\\\nonumber
&&=-\frac12\Omega^2_\perp\int_{-h}^{h}\pder{}z(u^z\rho) z^2\,dz=\\\nonumber
&&=\frac1{2}\Omega^2_\perp\int_{-h}^{h}\frac1r\pder{}r(r\rho u^r) z^2\,dz\\\nonumber
&&=-\frac{\dot M}{2\pi r}\Omega^2_\perp\der{}r\left(\frac1\Sigma\int_{0}^{h}\rho z^2\,dz\right)=\\\nonumber
&&=-\frac{\dot M}{2\pi r}\Omega^2_\perp\der{\eta_4}r.
\end{eqnarray}

\section{NUMERICAL METHOD}
\label{s.numerical}
\subsection{Vertical structure}
\label{s.numerical.vertical}

The set of ordinary differential equations describing the vertical
structure, i.e., Eqs.~(\ref{vs.dpdz}), (\ref{vs.dFdz})  and
(\ref{vs.gradient}) together with appropriate boundary conditions are
solved for a given BH spin on a three-dimensional grid spanned  by the radius $r$, the
central temperature $T_c$, and the advection  factor $f^{\rm adv}$. For
a given set of these parameters, we start the integration from the
equatorial plane ($z=0$), and the solution satisfying the outer
boundary condition is found as described at the end of \S\ref{s.verticaleq}.  The
resulting quantities describing the vertical structure 
($T_c$, $\Sigma$, $Q^{\rm adv}$, $P$, $\eta_1$, $\eta_2$, $\eta_3$,
and $\eta_4$),
together with $r$, are printed out to tables for subsequent
use in interpolation routines. As it turns out, the first two of these
parameters, $T_c$ and $\Sigma$, can be used to uniquely
determine all the other quantities characterizing the vertical structure,
including $f^{\rm adv}$ (see Fig.~\ref{f.scurve}).

Calculating the
full grid of vertical solutions for a single value of BH spin takes
about 5 hours on a 4-CPU workstation.

\subsection{Radial structure}
\label{s.numerical.radial}

By a series of algebraic manipulations of 
Eqs.~(\ref{eq_cont2}), (\ref{eq_ang6}),  (\ref{eq_rad3}), (\ref{e.dPdr}),
and (\ref{eq.qadv}),
 we obtain the following set of two ordinary differential equations for $V(r)$ and $T_c(r)$ 
\be
\frac{1}{1-V^2}\der{\ln V}{\ln r}=\frac{\cal N_{\rm 1}}{\cal D_{\rm 0}},
\label{eq_derV}
\ee
\begin{eqnarray}\nonumber
(4-3\beta)\der{\ln T_c}{\ln r}=\left(\frac{\cal N_{\rm 1}}{\cal D_{\rm 0}}+\frac{r(r-M)}\Delta\right)(\beta-\tilde\Gamma_1)&&\\\nonumber
-\frac{2\pi r^2}{\dot M\eta_3}\frac\Sigma P Q^{\rm adv}-(1-\beta)\derln{\eta_1}r-\beta\derln{\eta_2}r&&\\
-\derln{\eta_3}r-\frac{\Omega_\perp^2\Sigma}{P}\frac{\eta_4}{\eta_3}\derln{\eta_4}r
\label{eq_derT}
\end{eqnarray}
with $\cal N_{\rm 1}$ and $\cal D_{\rm 0}$ given by
\begin{eqnarray}\nonumber\label{e.NN}
{\cal N_{\rm 1}}&=&{\cal A}+\frac{2\pi r^2}{\dot M\eta_3}Q^{\rm adv}+
\frac{P}\Sigma
\left(\frac{r(r-M)}\Delta\tilde\Gamma_1+\right.\\
&+&\left.\derln{\eta_3}r\right)+\Omega^2_\perp\frac{\eta_4}{\eta_3}\derln{\eta_4}r
\end{eqnarray}
\begin{eqnarray}
{\cal D_{\rm 0}}&=&V^2-\tilde\Gamma_1\frac P\Sigma.
\end{eqnarray}
Typically, ${\cal D}_0$ vanishes close to the black hole, as ${\cal
  L}$ approaches ${\cal L}_{\rm in}$.

To obtain a solution one has to solve this system of two ordinary
differential equations, together with the following regularity
conditions at the sonic radius $r_S$, defined by the same conditions:
\be \left.{\cal N}\right|_{r_S}=\left.{\cal D_{\rm 0}}\right|_{r_S}=0,
\label{num_regcond}
\ee as well as outer boundary conditions given at some large radius
$r_{\rm out}$. The solution between the outer boundary and the
sonic point is found using the relaxation technique
\citep{numericalrecipes}, with ${\cal L}_{\rm in}$ treated as the
eigenvalue of the problem. 
The sonic point for $a_*=0$ is located at $5.9M$ for accretion rate $0.01\dot M_{Edd}$ and at $5.0M$
   for $2.0\dot M_{Edd}$.
To start the relaxation process we have to
provide a trial solution that is obtained by a method similar to the
one described in \cite{sadowski.slim} assuming the \cite{nt} outer
boundary conditions. Once the trial solution is found, one can start
the relaxation process with a free inner boundary corresponding to the
location of the sonic point. To find the solution inside the sonic
point we make a small step inward to cross the critical point and then
integrate down to BH horizon using a Runge-Kutta method of the fourth
order.
In this work we use 25 mesh
points spaced logarithmically in the radius on the section between the sonic point 
and $r_{\rm out}=1000M$. This particular number of grid
points is enough to resolve all disk features. We have verified that the results are accurately reproduced with a denser grid.

The parameters linked to the vertical structure ($P$, $Q^{\rm adv}$,
$\eta_1$, $\eta_2$, $\eta_3$, and $\eta_4$) for given $\Sigma$ and
$T_c$ are linearly interpolated from pre-calculated tables of the
vertical structure solutions
(for any value of $V$, $\Sigma$ is determined directly from mass conservation,
Eqs.~(\ref{masscon}), (\ref{eq_cont2})). 
The radial derivatives ${\rm d}\ln
\eta_1/{\rm d}\ln r$, ${\rm d}\ln \eta_2/{\rm d}\ln r$, ${\rm d}\ln
\eta_3/{\rm d}\ln r$, and ${\rm d}\ln \eta_4/{\rm d}\ln r$ are
evaluated numerically from the $\eta_1$, $\eta_2$, $\eta_3$, and
$\eta_4$ profiles in the previous iteration step. A relaxed solution
is obtained in a few iteration steps.  Once a solution outside the
sonic point is found we numerically estimate the radial derivatives of
$V$ and $T_c$ at the sonic point using values given at $r>r_S$ and use
these derivatives to start direct integration inside the sonic point.

The solution thus obtained may then be used as a trial solution when
looking for the relaxation solution of another slim disk, i.e., when
one of the three fundamental parameters ($\dot M$, $a$, $M$) has a
slightly different value.
Each relaxation step takes approximately 5
seconds on a single-CPU workstation.

\section{RESULTS}
\label{s.results}

In the following two sections we present and discuss both the
radial and vertical structure of slim accretion disks.  All the
solutions, if not stated otherwise, were computed assuming
$\alpha=0.01$ and $M=10~{\rm M_\odot}$.
\subsection{Disk radial structure}
\label{ss.radialstructure}

\begin{figure}
  \resizebox{\hsize}{!}{\includegraphics[angle=0]{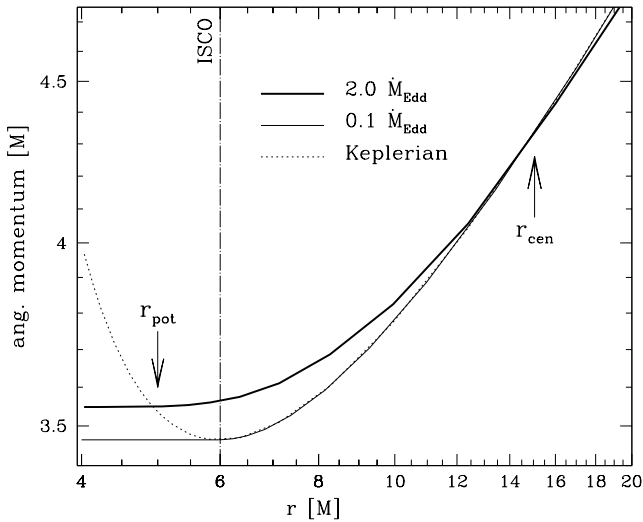}}
\caption{ Angular momentum ($u_{\phi}$) in a Schwarzschild slim disk for 
two accretion rates. For very low accretion rates the
angular momentum follows the Keplerian profile (dotted line) down to the ISCO. 
For high accretion rates the flow is
super-Keplerian between the ``center'' of the disk at $r_{\rm cen}$ 
and the ``potential spout''  at $r_{\rm pot}$. The vertical dot-dashed
line on this and subsequent figures denotes the location of the ISCO.}
  \label{f.angmom.a0}
\end{figure}

\subsubsection*{Angular momentum}

The angular momentum profiles for our accretion disk solutions near a
nonrotating BH are presented in Fig.~\ref{f.angmom.a0}. 
Results for two values of mass accretion rate are shown, $\dot M=0.1M_{\rm Edd}$
and $2.0M_{\rm Edd}$, where
$\mdot_{\rm Edd} = 16L_{\rm Edd}/c^2$ is the
critical accretion rate that for a disk around a nonrotating BH
approximately corresponds to the Eddington luminosity, $L_{\rm Edd}$.
 For the
lowest accretion rates the profiles follow the Keplerian profile and
reach its minimal value (${\cal L}_{\rm in}$ in Eq.~(\ref{eq_ang6})) at
the marginally stable orbit \citep[ISCO, ][]{bardeen-72}. The higher the accretion rate, the
stronger the deviation from the Keplerian profile. The disk is
sub-Keplerian at large distances and super-Keplerian at moderate
radii.  The Keplerian profile is crossed again at a point located
inside the marginally stable orbit, and corresponding to what is
usually called ``the cusp'' or ``the potential spout''.  For a
detailed study of the physics of the inner edge of a see
\cite{leavingtheisco}.

\subsubsection*{S-curves}

\begin{figure}
  \resizebox{\hsize}{!}{\includegraphics[angle=270]{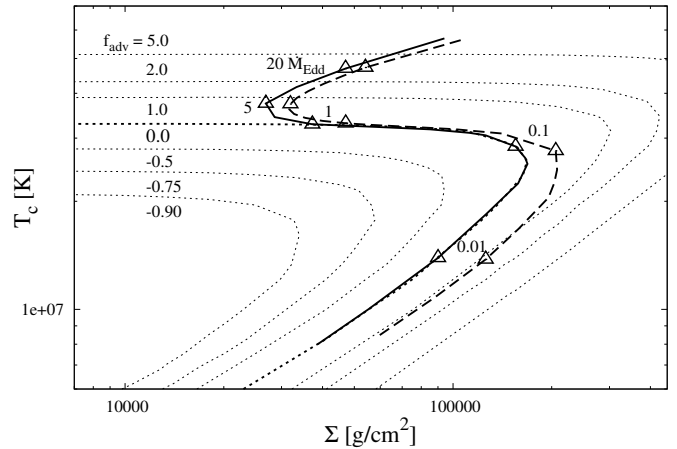}}
\caption{The $T_c$-$\Sigma$ plane at $r=20M$ for a nonrotating BH ($a_*=0$).
 The dotted lines connect
solutions for the vertical structure of slim disks that have the same value
of the advection parameter $f^{\rm adv}$. The locus of standard
(radiatively efficient, $f^{\rm adv}=0$)
 disk solutions is shown with the thick dotted line.
The solid thick line represents the vertical slim-disk solutions for
different accretion rates
(indicated by triangles), and the dashed line presents corresponding solutions
of the conventional polytropic slim-disk model
(see \S\ref{s.comparison}). The difference between the two lines in the low
$\dot M$ limit corresponds to the difference in $\Sigma$ between the two models
(see Fig.~\ref{f.3comp}).
}
  \label{f.scurve}
\end{figure}

Figure~\ref{f.scurve} presents slim disk
solutions at $r=20 M$ on the $T_c$-$\Sigma$ plane, for a nonrotating BH. Solutions of the 
polytropic, height-averaged models are presented for comparison;
for detailed discussion see \S\ref{s.comparison}. The locus of solutions
for various values of the mass accretion rate has the shape of the
so-called ``S-curve'' \citep{slim}.  The lower, gas-pressure dominated
branch accurately follows the track of radiatively efficient solutions
($f^{\rm adv}=0$).  The middle, radiation-pressure dominated branch is
reached at $\mdot\approx0.1\mdot_{\rm Edd}$. As advection becomes
significant, the slim-disk solution leaves the $f^{\rm adv}=0$ track
and moves to higher advection rates (S-curve). Around
$\mdot=5\mdot_{\rm Edd}$, the solutions enter the upper advection-dominated branch corresponding to $f^{\rm adv}>1.0$  (more than 50\%
of heat stored in the accreted gas).  At $\mdot=20\mdot_{\rm Edd}$
this rate almost increases up to 80\% ($f^{\rm adv}\approx4.0$).

\subsubsection*{Surface density}

\begin{figure}
\centering
 \subfigure
{
\includegraphics[height=.35\textwidth]{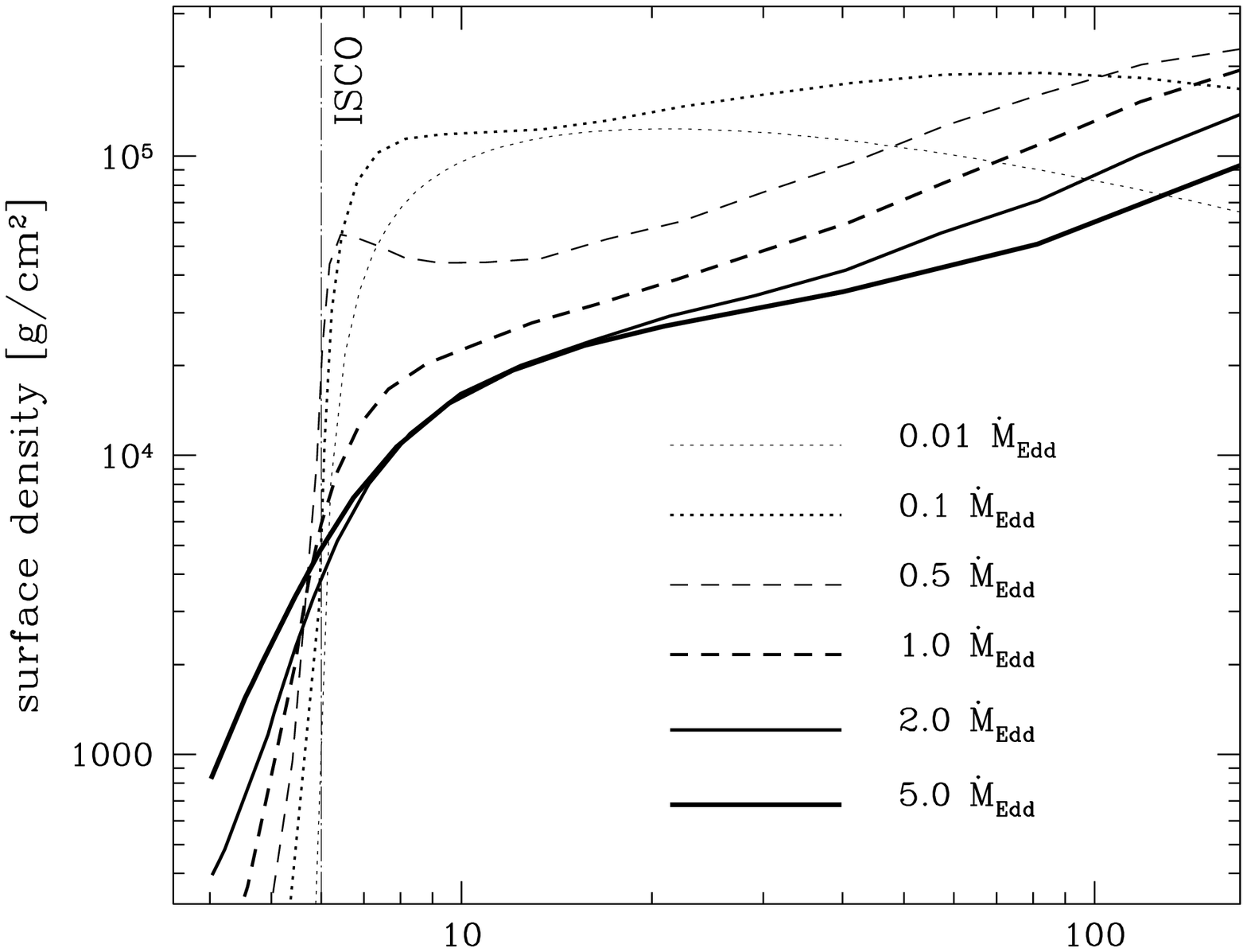}
}
\\
\vspace{-.07\textwidth}
 \subfigure
{
\includegraphics[height=.35\textwidth]{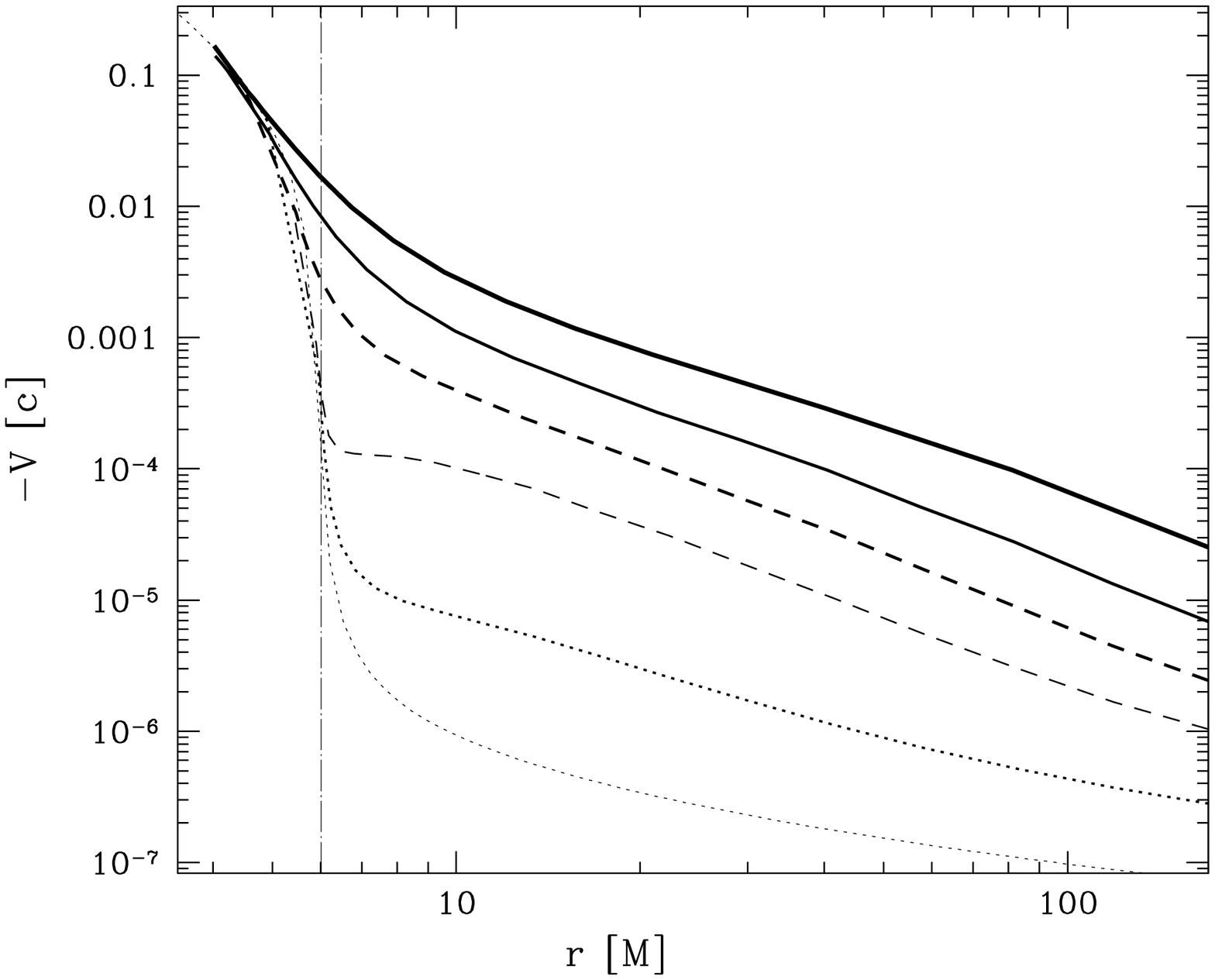}
}
\caption{ Profiles of the surface density (upper panel)
and corresponding values of radial velocity $V$ (bottom
panel) of a slim disk for
a nonrotating BH. Solutions for different accretion rates
are presented.}
\label{f.surfdens.a0}
\end{figure}

Profiles of the surface density for 2-D slim-disk solutions for
a non-rotating BH are presented in the upper panel of Fig.~\ref{f.surfdens.a0}. 
Different 
regimes, corresponding to different branches of the ``S-curve''
on the ($\Sigma$, $T_c$) plane are visible.
For large radii the surface density increases with increasing 
accretion rate (the lower gas-pressure dominated branch), while this relation is opposite for
moderate radii (the middle radiation-pressure
dominated branch). For accretion rates $\mdot>5.0\mdot_{\rm Edd}$ the 
upper advection-dominated branch would be reached. The local maxima
in the surface density profiles \citep[discussed in detail
 in, e.g.,][]{sadowski.slim}
 are visible for moderate accretion rates ($\sim 0.5\mdot_{\rm Edd}$).
The bottom panel of Fig.~\ref{f.surfdens.a0} presents corresponding 
profiles of the radial velocity $V$ as measured by an observer corotating
with the fluid. 

\begin{figure}
  \resizebox{\hsize}{!}{\includegraphics[angle=0]{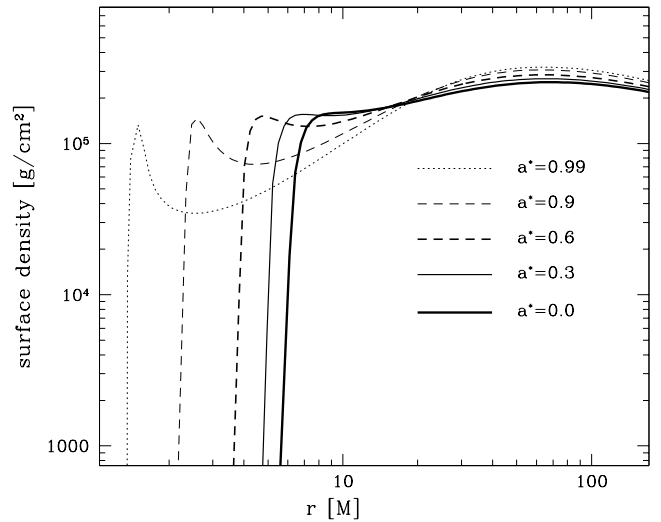}}
\caption{ Profiles of the surface density for slim disks 
at a constant accretion rate ($\dot M = 0.1\dot M_{\rm Edd}$) 
and various BH spins. }
  \label{f.surfdens.aN}
\end{figure}

The surface density dependence on BH rotation is presented in 
Fig.~\ref{f.surfdens.aN}. The profiles are shifted to lower radii as the inner 
edge of the disk moves inward for higher BH spins. The outer
parts of the accretion disk are insensitive to the metric.



\subsubsection*{Optical depth}

\begin{figure}
\centering
 \subfigure
{
\includegraphics[height=.35\textwidth]{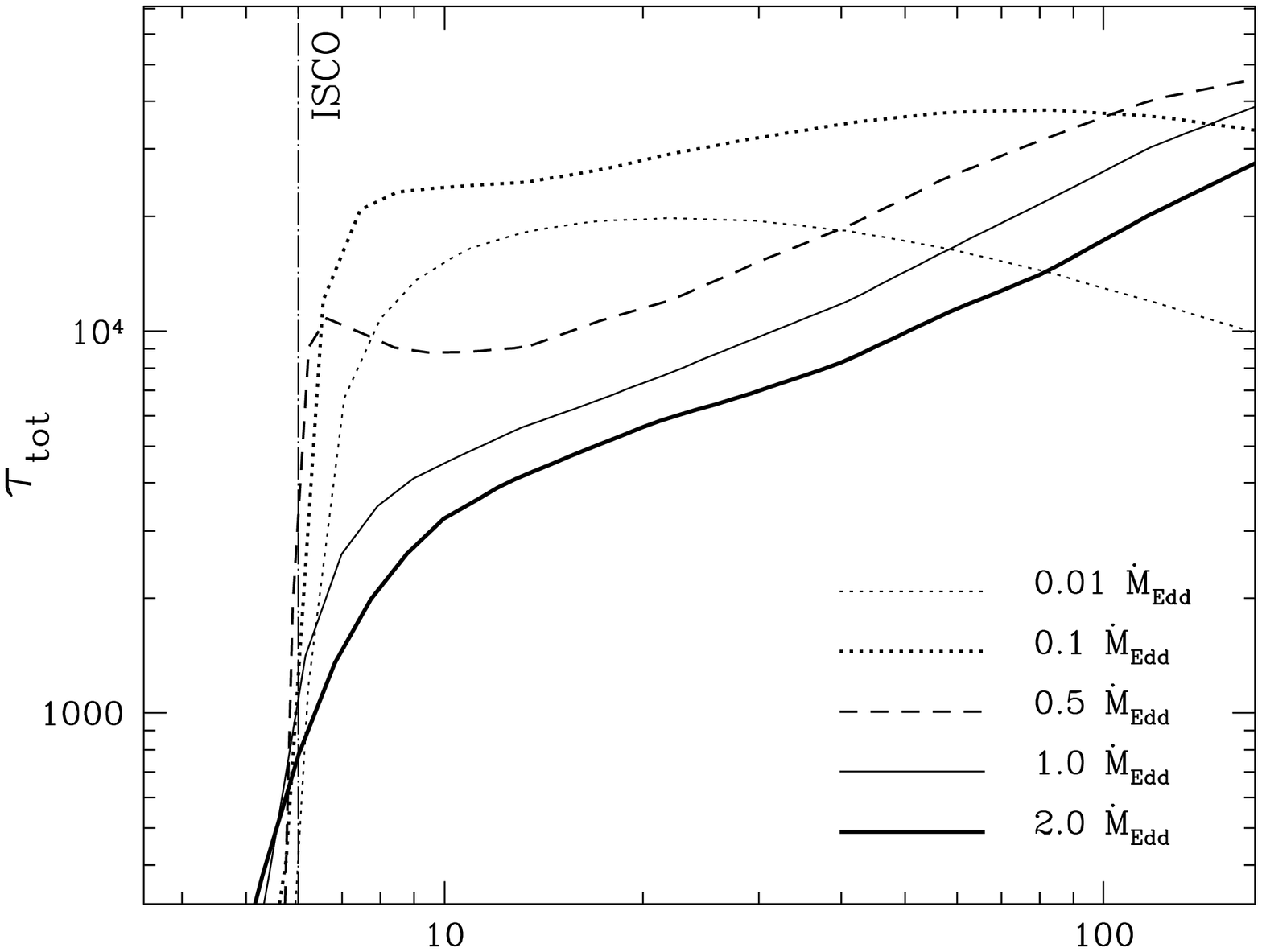}
}
\\
\vspace{-.07\textwidth}
 \subfigure
{
\includegraphics[height=.35\textwidth]{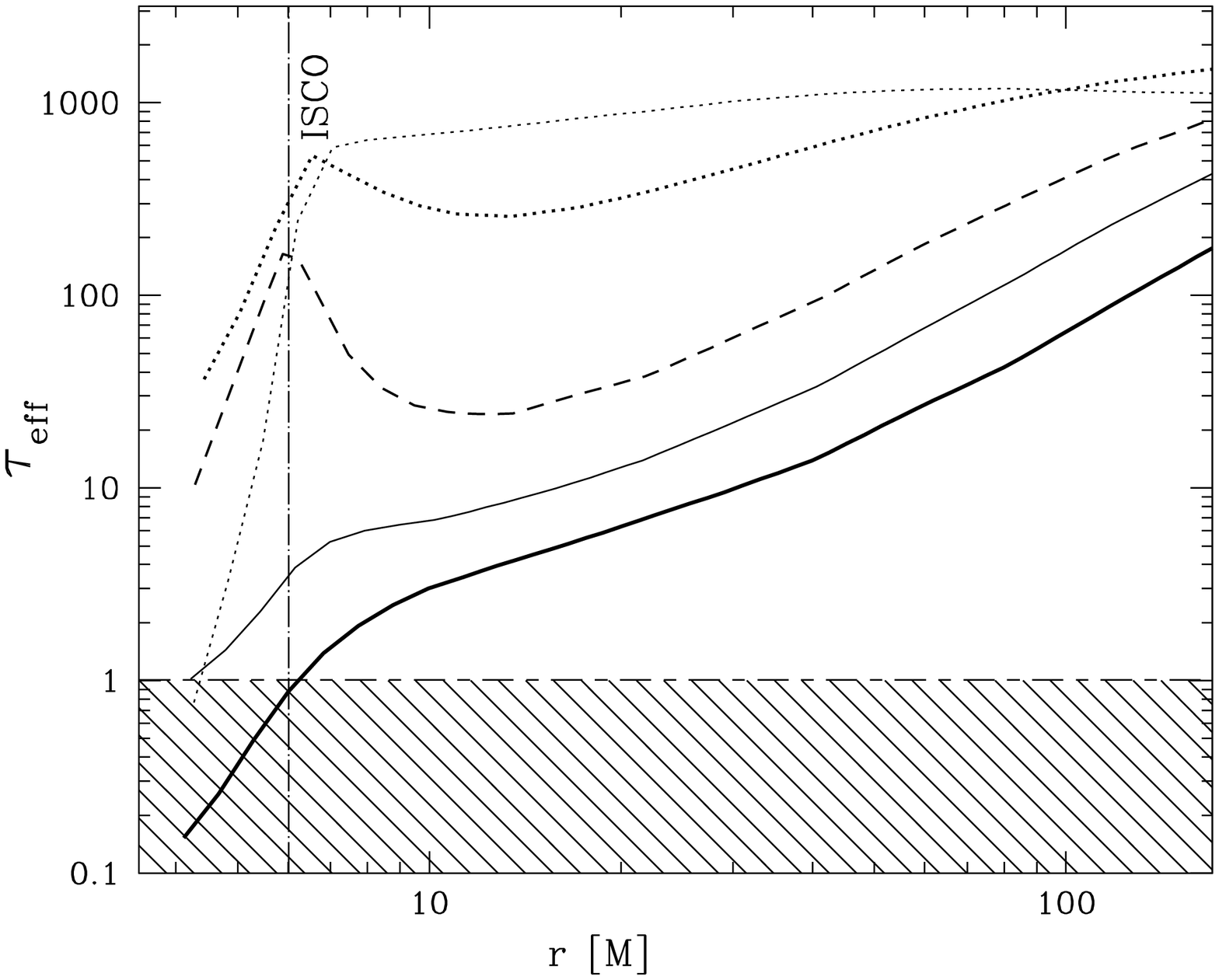}
}
\caption {Optical depth for $\alpha=0.01$ slim disks
around a  Schwarzschild black hole at different accretion rates. 
{\sl Top panel:} The total optical depth as a function of the radius.
{\sl Bottom panel:} The effective optical depth as a function of the radius. 
The ISCO is shown at $r=6M$. The shaded region in the bottom plot
indicates the region where the diffusive approximation is invalid.
}
 \label{f.optdepth.a0}
\end{figure}

In the top panel of Fig.~\ref{f.optdepth.a0} we plot the total
optical depth of the vertical slim disk solutions for different accretion 
rates ($a_*=0$). The total optical depth
$$\tau_{tot}=\int_0^h\kappa_R\rho~{\rm dz},$$
where the total opacity coefficient $\kappa_R$,
which includes the processes of absorption and scattering,
is closely related to the surface density.
Indeed, the radial profiles of the optical depth 
shown in Fig.~\ref{f.optdepth.a0} follow the corresponding profiles of 
surface density. Any differences in the profiles
come from the dependence of the opacity coefficient 
on local density and temperature. Outside the ISCO the total
optical depth is always large ($\tau_{tot}>10^3$).

%
%

\begin{figure}
 \centering \includegraphics[angle=0,width=.45\textwidth]{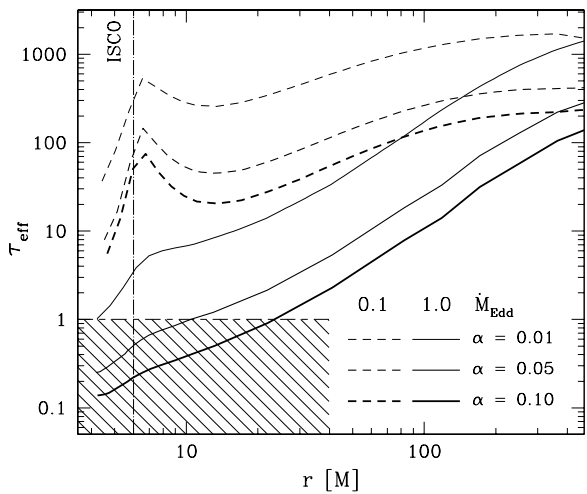}
\caption{ Profiles of the effective optical depth of a Schwarzschild
slim disk for three values of viscosity ($\alpha=0.01$, $0.05$, and $0.1$),
calculated for two accretion rates, $0.1\mdot_{\rm Edd}$ (dashed lines)
and $1.0 \mdot_{\rm Edd}$ (solid lines). }
  \label{f.1alpha}
\end{figure}

The diffusive approximation for radiative transport may only be used if photons are absorbed, otherwise LTE cannot be established. In a scattering-dominated atmosphere, the effective optical depth is then the relevant quantity to be used in checking for the self-consistency of the diffusive approximation. 
The bottom panel of Fig.~\ref{f.optdepth.a0} presents corresponding profiles
of the effective optical depth, which is
 estimated in the following way:
$$\tau_{\rm eff}=\int_0^h\sqrt{(\kappa_R-\kappa_{es})\kappa_R}~\rho\,{\rm dz}$$
where $\kappa_{es}=0.34~{\rm cm^2 g^{-1}}$ 
is the mean opacity for Thomson electron scattering.
For $\dot M>0.3\dot M_{\rm Edd}$ the inner region of the disk 
becomes radiation-pressure dominated 
(it enters the middle branch on the corresponding S-curve),
 the surface density decreases with increasing
accretion rate, and electron scattering begins to dominate
absorption. Therefore, the effective optical depth decreases with
increasing
accretion rate and reaches values  $\tau_{\rm eff}<1$ 
in the inner parts of the disk for accretion rates above $1.0\dot M_{\rm Edd}$. 
As a result, for $\dot M>1.0\dot M_{\rm Edd}$, 
the diffusive approximation can no longer be applied, 
and our present approach to solving for the disk structure breaks down.
 In Fig.~\ref{f.1alpha}
we exhibit the dependence of the effective optical depth on the value
of $\alpha$.
In general, $\tau_{\rm eff}$ is inversely proportional to  $\alpha$
(as is the surface density). At lower accretion rates ($\dot M<0.1\mdot_{\rm Edd}$) 
the effective optical depth remains large even for high values of $\alpha$.

\subsubsection*{Flux profiles}
\begin{figure}
  \resizebox{\hsize}{!}{\includegraphics[angle=0]{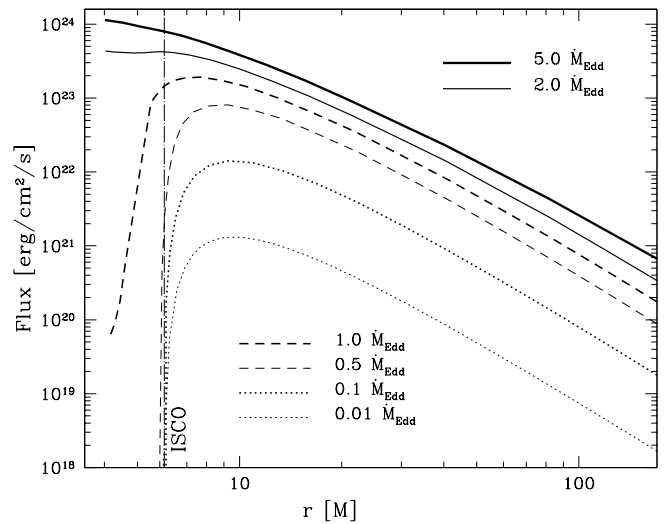}}
\caption{Flux emitted from the surface of a slim disk at five
accretion rates onto a Schwarzschild black hole.
At high accretion rates significant emmision from within the
ISCO is clearly visible in the figure.}
  \label{f.flux.a0}
\end{figure}

Profiles of the flux emitted from the disk surface ($F=\sigma
T_{\rm eff}^4$)  in the case of a nonrotating BH are presented in
Fig.~\ref{f.flux.a0}. Results corresponding to accretion rates
from $0.01$ up to $5.0\mdot_{\rm Edd}$ are shown. For the lowest
rates the emission from inside the marginally stable orbit is
negligible as expected in the standard accretion disk
models. This is no longer true for higher accretion rates,
and the advection of energy causes  significant emission from smaller
radii \citep{leavingtheisco}. For super-Eddington accretion rates the
emitted flux continues to grow with a decreasing radius even inside the
marginally stable orbit. Radiation coming from the direct vicinity of the black hole is
suppressed by the gravitational redshift (the $g$-factor).
Therefore, an observer at infinity will observe a
maximum in the profile of the effective temperature even for
the highest accretion rates.

\begin{figure}
  \resizebox{\hsize}{!}{\includegraphics[angle=0]{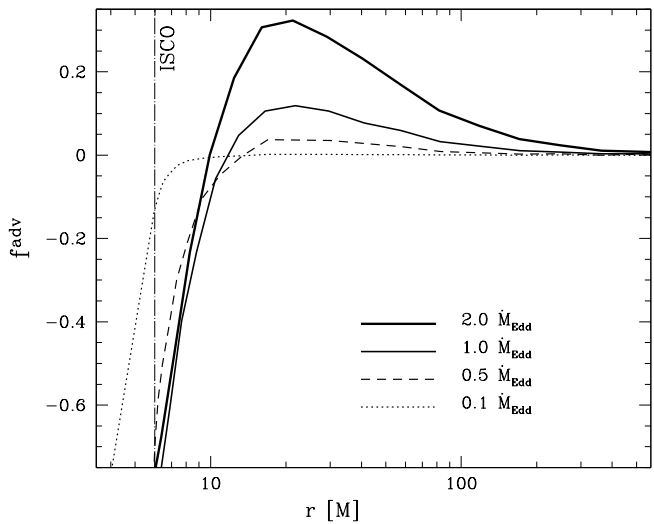}}
\caption{ Profiles of the advection coefficient $f^{\rm adv}$
for different accretion rates (Schwarzschild black hole).}
  \label{f.fadv.a0}
\end{figure}

The increase in the advective flux with increasing accretion rate
is clearly visible in Fig.~\ref{f.fadv.a0}. The ratio of the
heat advected to the amount of energy emitted 
is presented for different accretion rates. For very low accretion rates 
these profiles approach the limit of a radiatively efficient disk
($f^{\rm adv}\equiv 0$), and the advection component becomes significant for higher accretion rates.
Some part (up to $30\%$ at $r=20M$ for $2.0\mdot_{\rm Edd}$)
of the energy generated at moderate radii is advected with matter and radiated 
away at $r<10M$. This
causes the significant change in the emitted flux profile
at the higher accretion rates visible in 
Fig.~\ref{f.flux.a0}.

\begin{figure}
  \resizebox{\hsize}{!}{\includegraphics[angle=0]{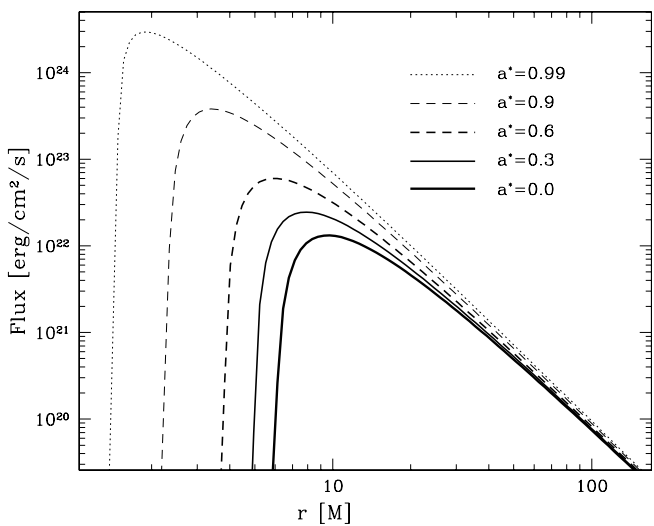}}
\caption{ Flux profiles at fixed accretion rate ($\dot M=0.1\dot M_{\rm Edd}$)
for five values of BH spin.}
  \label{f.flux.aN}
\end{figure}

In Fig.~\ref{f.flux.aN}  we present the emitted flux profiles for
different BH angular momenta at a constant accretion rate
$\mdot=0.1\mdot_{\rm Edd}$. These profiles coincide at large radii
where the influence of the BH  rotation is negligible; however, the
higher the BH spin, the closer to  the horizon the marginally stable
orbit. Therefore, in  the case of rotating BHs, the accreting matter
can move much deeper into the gravity well, compared to nonrotating
BHs. This effect leads to an increase in the disk luminosity and
hardening of its spectrum, which can be inferred from
Fig.~\ref{f.flux.aN}---the higher the spin, the higher the disk luminosity,
 and the
higher the flux (which corresponds to the effective temperature).

\begin{figure}
  \resizebox{\hsize}{!}{\includegraphics[angle=0]{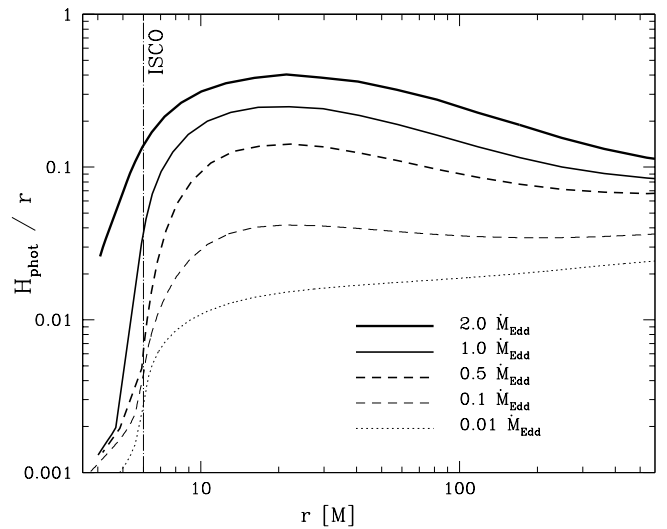}}
\caption{ The height of the photosphere at different accretion rates
onto a Schwarzschild black hole. }
  \label{f.Hphot.a0}
\end{figure}

\begin{figure}
  \resizebox{\hsize}{!}{\includegraphics[angle=0]{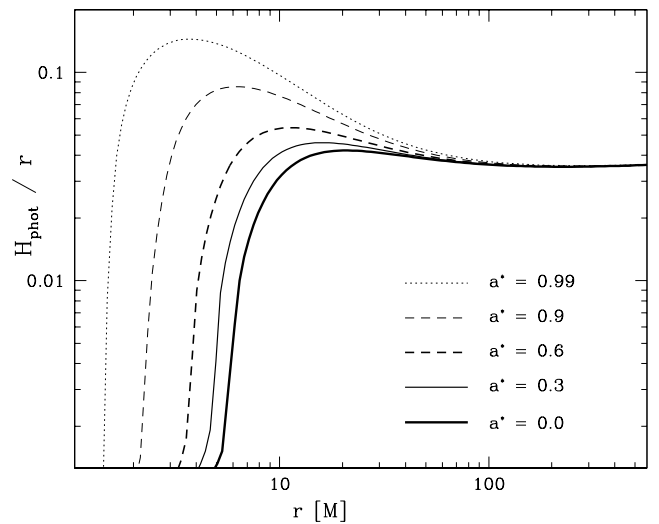}}
\caption{Profiles of photospheric height at a constant accretion rate
($\dot M=0.1\dot M_{\rm Edd}$) and various BH spins. }
  \label{f.Hphot.aN}
\end{figure}

\subsubsection*{Photosphere location}

The flux observed at infinity may be obtained by performing ray tracing
of photons emitted from the accretion disk (see \S\ref{s.comparison}).
Scattering in the layer
above the photosphere must also be taken into account.
An accurate calculation requires detailed knowledge of
atmospheric properties, including the location of the photosphere.
This is particularly
important when accretion rates are high and the disk is no longer geometrically
thin \citep{sadowski.photosphere}. 

\begin{figure}
\centering
\resizebox{\hsize}{!}{\includegraphics[width=.7\textwidth]{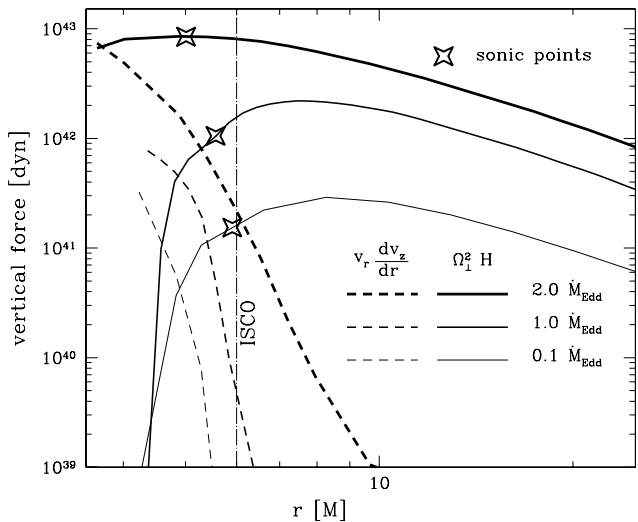}}
\caption{Comparison of the dynamical ($v_r{\rm d}v_z/{\rm d}r\approx V{\rm d}(V{\rm d}H/{\rm d}r)/{\rm d}r$) and gravitational ($\Omega_\perp^2 H$) components
of the vertical equilibrium equation at the photosphere. The stars denote locations of the sonic
radii.}
\label{f.vertbal}
\end{figure}

In Fig.~\ref{f.Hphot.a0} we plot the profiles for the
$z$-location of the photosphere, $H_{\rm phot}$, obtained in our
model at different accretion rates for a nonrotating black hole.
Clearly, for high accretion rates, $\mdot>0.1\mdot_{\rm Edd}$,
 the inner regions become thicker (effects of
radiation pressure). For $\mdot = 1.0\mdot_{\rm Edd}$,
 the ratio $H_{\rm phot}/r$ reaches a value as high as $0.25$. 
Close to the ISCO the height of the photosphere rapidly
  decreases because of vigorous cooling (compare Fig.~\ref{f.fadv.a0}).
Although the rapid change in disk thickness violates the assumption of the hydrostatic equilibrium that we make when solving for the disk vertical structure, the accelerations connected with the vertical motions
involved are much lower than the vertical component of gravity.
Indeed, the vertical accelerations are close to
$v_r dv_z/dr\sim v_rd(v_rdH/dr)/dr\sim v_s^2 d(H/r)/dr\sim r\Omega_\perp^2 (H/r)^2\sim\Omega_\perp^2 H (H/r) $.
In deriving this estimate we liberally assumed that  $dH/dr\sim 1$ and
used the fact that the rapid decrease in disk height occurs near the sonic point, 
while the speed of sound $v_s$ is approximately $r\Omega_\perp (H/r)$.
Thus the acceleration terms modifying Eq.~(\ref{vs.dpdz}) would be
smaller than
the gravitational acceleration by a factor of a few percent: $(H/r)\sim10^{-1}$.
In Fig.~\ref{f.vertbal} we plot the dynamical and gravitational components 
of the vertical equilibrium equation at the photosphere. It is clear that the former
is at least $10$ times smaller than the latter at the sonic radii for $\dot M\le\dot M_{\rm Edd}$.
Therefore, in all likelihood, our results correctly describe the disk structure that would be obtained without assuming strict hydrostatic equilibrium. 

In Fig.~\ref{f.Hphot.aN} we present radial photosphere profiles,
at a fixed accretion rate and different
values of the BH spin. The photospheric heights
coincide for large radii.  In the inner regions of the disk, the height of the
photosphere increases with BH spin, reflecting the increased
luminosity and radiation pressure.

\subsection{Vertical structure}
\label{ss.vertstructure}

\begin{figure}
\centering
\resizebox{\hsize}{!}{\includegraphics[width=.7\textwidth]{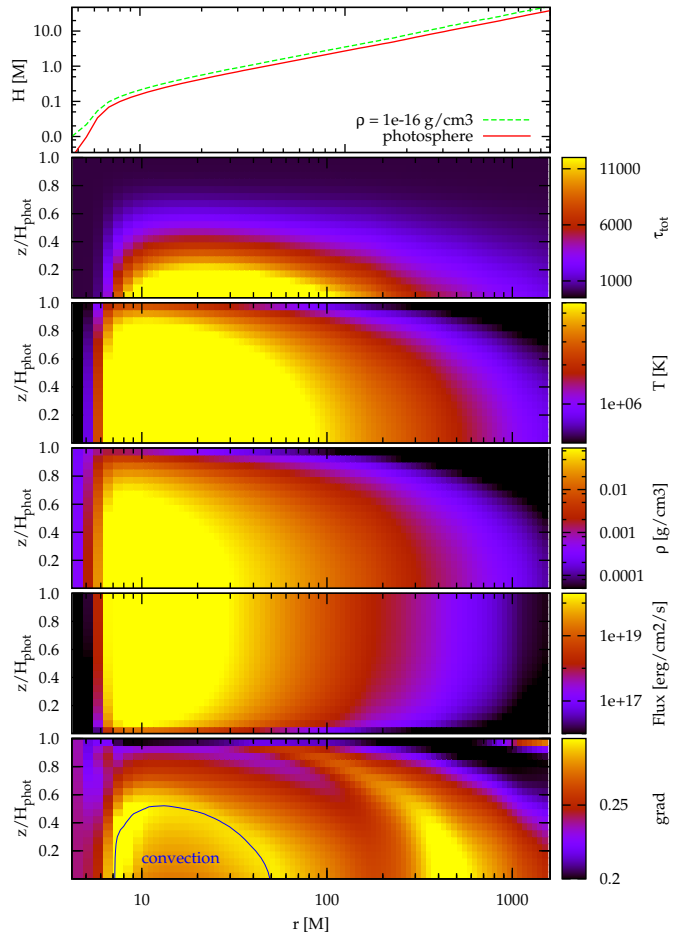}}
\caption{Vertical structure of a slim disk
for $\mdot=0.01\dot M_{\rm Edd}$ and $a_*=0$. 
The top panel presents the surface of the disk (green dashed line),
and the photospheric surface (red solid line). 
The other panels present the structure of the disk below the 
photosphere. {\sl Top to bottom:} total optical thickness, temperature, density,
vertical flux of energy, and the termodynamical gradient.
 The blue solid line in the bottom panel delimits
the convective region.}
\label{f.rv.md0.01.a0}
\end{figure}

In Fig.~\ref{f.rv.md0.01.a0} we present a
vertical cross-section of a Schwarzschild slim disk
for $\mdot=0.01\mdot_{\rm Edd}$.
At this accretion rate the disk is radiatively efficient and no advection
of entropy is expected. The top panel presents the radial profiles
of the photosphere and the disk surface (defined as a layer
with $\rho=10^{-16}\,$g/cm$^3$).

\begin{figure}
\centering
\resizebox{\hsize}{!}{\includegraphics[width=.7\textwidth]{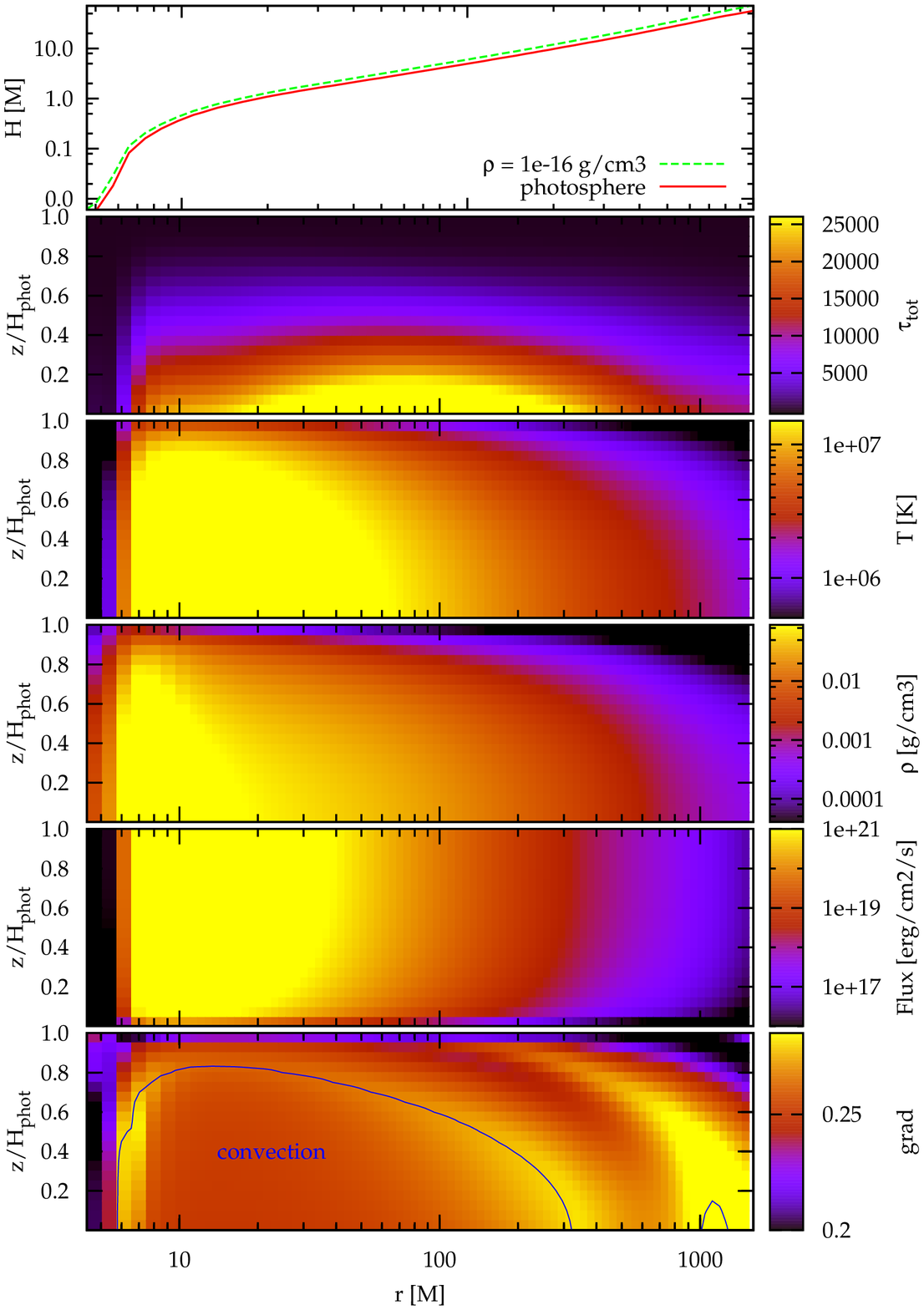}}
\caption{Same as Fig.~\ref{f.rv.md0.01.a0} but for 
$\mdot=0.1\dot M_{\rm Edd}$.}
\label{f.rv.md0.1.a0}
\end{figure}
\begin{figure}
\centering
\resizebox{\hsize}{!}{\includegraphics[width=.7\textwidth]{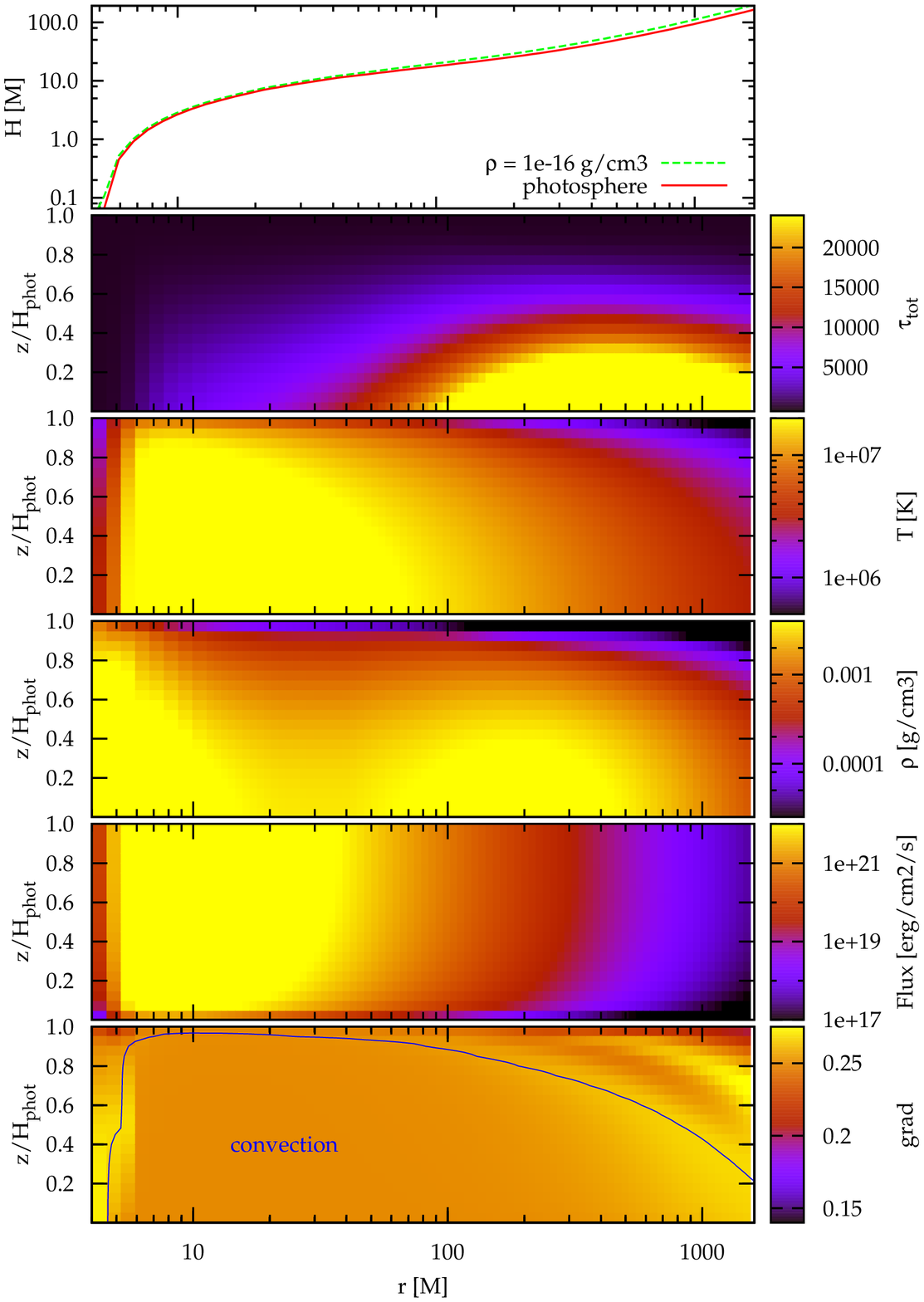}}
\caption{Same as Fig.~\ref{f.rv.md0.01.a0} but for 
$\mdot=1.0\dot M_{\rm Edd}$.}
\label{f.rv.md1.0.a0}
\end{figure}

 The total optical depth reaches
values as high as $\sim20000$ on the equatorial plane and decreases
monotonically towards the photosphere at $\tau=2/3$
(Fig.~\ref{f.rv.md0.01.a0}, second panel from the top).
Within the marginally stable orbit, the total optical depth
significantly decreases, as shown in Fig.~\ref{f.optdepth.a0}.

The third panel of Fig.~\ref{f.rv.md0.01.a0} presents the
temperature, which decreases with height from $T_c$ at the equatorial plane
to $T_{\rm eff}$ at the photosphere. The maximum value of the temperature is 
attained on the equatorial plane at $r\sim10M$.

Density is presented in the fourth panel,
and the next panel presents the vertical flux that is generated
inside the disk according to Eq.~\ref{vs.dFdz}. It is set to zero on
the equatorial plane by reflection symmetry, and then rapidly increases,
because in an alpha disk the dissipation is proportional to the pressure,
which reaches its maximum on the $z=0$ plane. Close to the disk
surface, where  pressure is almost negligible, the flux slowly settles
down to the emitted value. At the accretion rate chosen for the figure the flux
rapidly decreases inside the marginally stable orbit.

The bottom panel of Fig.~\ref{f.rv.md0.01.a0} presents the nonmonotonic 
distribution of the termodynamical gradient
(Eq.~(\ref{vs.gradient})), which ranges between
$0.2$ and $0.4$. Therefore, the disk's vertical structure cannot be
described by a simple polytropic relation. Moreover, in the region of the
highest temperature
(close to the equatorial plane at moderate radii, $r\approx 20M$),
the heat is transported upward through convection.

The vertical structure of an accretion disk with ten times higher accretion
rate, $\mdot=0.1\mdot_{\rm Edd}$, 
is presented in Fig.~\ref{f.rv.md0.1.a0}. The
general picture remains the same, since the advection of heat is still 
insignificant. However, as the temperatures increase,
 the inner disk regions become dominated by radiation pressure. 
 For $0.1\mdot_{\rm Edd}$ the 
convective region extends from the marginally stable orbit up to $300M$ and
covers more than half of the disk thickness.


The disk structure is significantly different in the case of
a high accretion rate (e.g., $1.0\mdot_{\rm Edd}$), with a 
significant amount of advection. The 
vertical cross-sections of the slim disk  are presented in 
Fig.~\ref{f.rv.md1.0.a0}. The inner regions are dominated by radiation 
pressure, so the disk geometrically
 thickens and the photosphere is now higher.
The total optical depth mostly follows the surface density
 (compare Fig.~\ref{f.optdepth.a0}), and therefore decreases 
considerably  towards the black hole in the inner parts of the disk.

The temperature maximum is again
 located at the equatorial plane close to $r\approx 10M$. 
The maximum of the effective temperature 
(corresponding to the vertical flux shown
in the fifth panel (compare also Fig.~\ref{f.flux.a0}) 
is shifted inwards, down to $r\approx 8M$.

The fourth panel of Fig.~\ref{f.rv.md1.0.a0} presents the density distribution.
Despite the fact that the surface density monotonically increases
outwards (see Fig.~\ref{f.surfdens.a0}), $\rho$ has
 two maxima in the equatorial plane: at $r\approx200M$
 and $r\approx 6M$. Finally, the bottom panel presents the termodynamical gradient distribution.
 For such a high accretion rate, the convective zone extends
nearly to the photosphere for $r<200M$ and is present up to $r=2000M$.

\begin{figure}
\centering
\resizebox{\hsize}{!}{\includegraphics[angle=270,width=.7\textwidth]{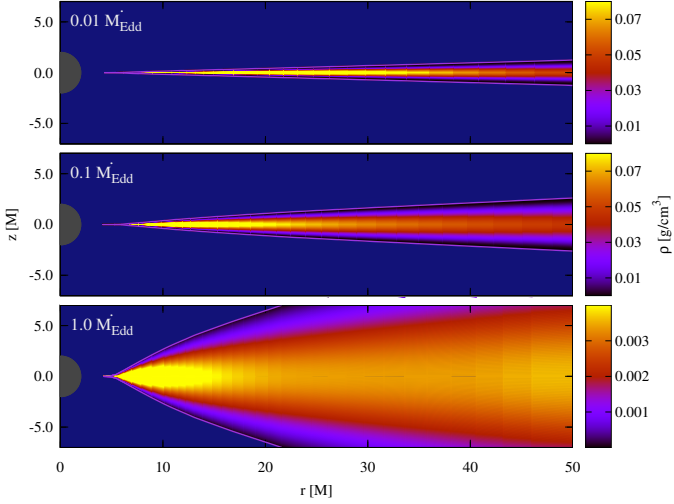}}
\caption{Meridional profiles of density for three accretion rates: $
\mdot=0.01$ (top), $0.1$ (middle) and $1.0 \mdot_{\rm Edd}$ (bottom
panel) in (r,z) coordinates. The violet boundaries show the location of
the photosphere. The black hole is described by
$M_{\rm BH}=10\msun$, and $a_*=0$.}
\label{f.slice}
\end{figure}

In Fig.~\ref{f.slice} we present the density in the meridional plane
for three different accretion rates. The
equatorial plane lies in the middle of each plot. The violet boundaries
denote the photosphere. The two maxima of density at the equatorial plane
for $\mdot=1.0\mdot_{\rm Edd}$) are clearly visible also in this representation,
as are some other features discussed above.

\section{Comparison with height-averaged slim disk solutions}
\label{s.comparison}

In this section we compare our 2-D slim disk solutions,
 where the radial structure
equations are coupled to those for the vertical structure, with the
standard polytropic slim disk model, in which the slim-disk equations and properties are
averaged over the thickness of the disk \citep[e.g.,][]{KatoBook}. 
We note here that the slim disk solutions presented in one
of our previous papers \citep{sadowski.slim} did not follow
the polytropic formalism, assuming different relations between
the vertically integrated quantities and their values on the
equatorial plane (e.g., $\Sigma=2\rho_0 H$ instead of $\Sigma=2I_N\rho_0H$, where $I_N$ is defined in Eq.~(\ref{e.IN})) In this work, we also use a more general form of
the energy equation (compare our Eq.~(\ref{eq.qadv}) with  
Eq.~(6) in \cite{sadowski.slim}).


Let us assume the polytropic equation of state with the polytropic
index $N$: $p=K\rho^{1+1/N}$. The vertical integration of the hydrostatic 
equilibrium formula, Eq.~(\ref{vs.dpdz}), gives
\be
\rho=\rho_0\left(1-\frac{z^2}{H^2}\right)^N.
\label{rho.polytropic}
\ee
We also assume 
\be
T=T_0\left(1-\frac{z^2}{H^2}\right).
\ee
One can now calculate analytical
formulae for $\eta_1$ to $\eta_4$  (Sect.~\ref{s.radialeq}):
\begin{eqnarray}\nonumber
\eta_1&=&\frac1{T_0^4}\int_0^HT^4\,dz=I_4 H\\\nonumber
\eta_2&=&\frac2{\Sigma T_0}\int_0^H\rho T\,dz=I_{N+1}/I_N\\\nonumber
\eta_3&=&\frac1P\left(\frac1{\gamma-1}\frac{k}\mu\frac{I_{N+1}}{I_N}\Sigma T_C+2I_4aT^4_CH\right)\\\nonumber
\eta_4&=&\frac1{\Sigma}\int_0^H\rho z^2\,dz=J_N H^2,
\end{eqnarray}
where 
\begin{equation}
I_N=\frac{\sqrt{\pi}}2\frac{\Gamma(1+N)}{\Gamma(3/2+N)}\overset{N\in\mathbb{N}}{=}\frac{(2^NN!)^2}{(2N+1)!}\\
\label{e.IN}
\end{equation}
\be
J_N=\frac14\frac{\Gamma(3/2+N)}{\Gamma(5/2+N)}=\frac1{6+4N}.
\ee

In this approach we do not solve the vertical structure consistently,
 so
we need to make some additional assumptions about the vertical
equilibrium of forces and radiation transfer. Following
other authors, we simplify the hydrostatic equilibrium (Eq.~(\ref{vs.dpdz}))
by applying a finite difference approximation and write
\be
H^2\Omega_\perp^2=(2N+3)\frac P\Sigma.
\ee
One has to remember that disk thickness, $H$, defined in this way
is not the exact location of the photosphere. Therefore we introduce a
factor $f_H$ relating these quantities,
\be
H_{\rm phot}=f_H H.
\ee
Assuming that radiation is transported in the vertical direction
through diffusion, the radiative flux is given by
\begin{equation}
 {\cal F}(z)=-\frac{16\sigma T^3}{3\kappa\rho}\der Tz.
\end{equation}
Under the one-zone approximation one obtains the following formula
for the total flux emitted from disk surface,
\be
F=f_F\frac{64\sigma T_C^4}{3\Sigma\kappa},
\ee
where factor $f_F$ has been introduced to account for inaccuracies 
arising from this approximation, as well as from the dominance
of the disk convection in certain regions (as discussed in \S\ref{ss.vertstructure}). 
Now, advective cooling takes the form
\be
Q^{\rm adv}=f^{\rm adv}F=-\alpha P\frac{A\gamma^2}{r^3}\der\Omega r - f_F\frac{64\sigma T_C^4}{3\Sigma\kappa}.
\ee

Usually, the following values of $N$, $f_H$ and $f_F$ are assumed:
\begin{eqnarray}\nonumber\label{poli.param}
N&=&3.0\\
f_H&=&1.0\\\nonumber
f_F&=&1.0 
\end{eqnarray}

\begin{figure}
 \centering \includegraphics[angle=0,width=.45\textwidth]{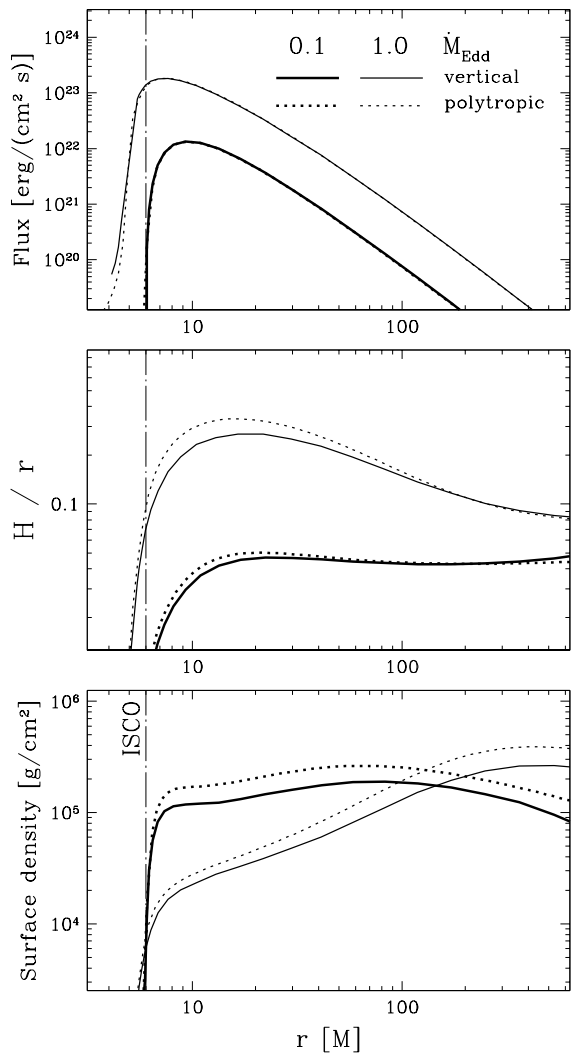}
\caption{ Comparison of the flux, disk thickness (the height of the photosphere for models presented in this paper and of the zero-density surface for 1-D polytropic models)
 and surface
density profiles calculated using the 2-D (this paper, solid lines)
 and the usual polytropic ($N=3$, dotted lines)
slim disk models for $\alpha=0.1$.
The solutions for two accretion rates ($0.1$ and $1.0 \mdot_{\rm Edd}$)
are presented with thick and thin lines, respectively.}
  \label{f.3comp}
\end{figure}

In Fig.~\ref{f.3comp} we compare radial profiles of the flux, photospheric
height and surface density of the 2-D
slim disk model described in this paper (consistently
taking its vertical structure into account)
with profiles obtained for the polytropic, height-averaged, slim disk, 
as presented in this section with parameters defined in Eq.~(\ref{poli.param}).
The comparison was carried out for two accretion rates
($0.1$ and $1.0\mdot_{\rm Edd}$), with $\alpha=0.01$. 
 For the lower accretion rate ($0.1\mdot_{\rm Edd}$), the disk is 
radiatively efficient (advective cooling is negligible), and therefore 
both flux profiles almost coincide and correspond to the
Novikov \& Thorne solutions. 
However, the 
disk photosphere location in the polytropic model turns out
to be overestimated by more than 20\% in the region corresponding
to the maximal emission.
 The profiles of the surface density do not coincide either,
the 2-D solutions giving values
 $\sim25\%$ lower than the corresponding polytropic solutions
(compare Fig.~\ref{f.scurve}).

For the higher accretion rate ($1.0 \mdot_{\rm Edd}$),
the flux profiles remain similar (up to 1\%).
However,  as advection becomes important, the emission is shifted
inwards with respect to the Novikov \& Thorne profile.
The photosphere location in the polytropic model is overestimated
by $\sim30\%$ 
and the surface density by $\sim20\%$.

\begin{figure}
 \centering \includegraphics[angle=0,width=.45\textwidth]{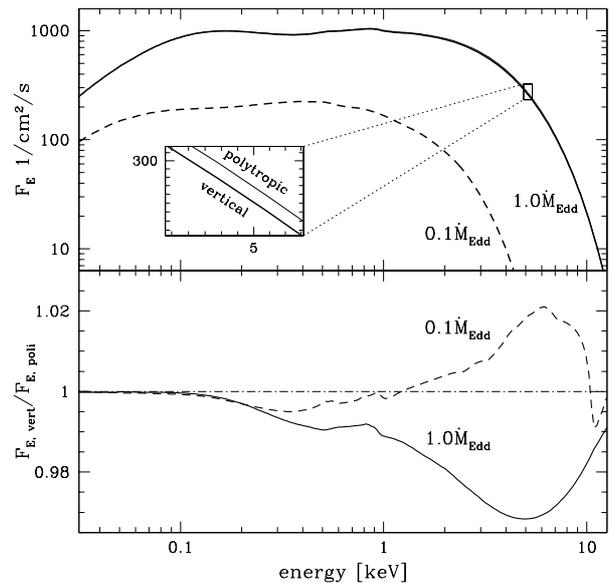}
\caption{ 
The upper panel presents spectral profiles of the 2-D and polytropic
solutions for two accretion rates 
($0.1$ and $1.0 \mdot_{\rm Edd}$), at inclination angle 
$i=70^o$ and distance $d=10{\rm kpc}$.
The bottom panel presents ratios of the corresponding spectra
(of 2-D to polytropic solutions)
for both accretion rates.}
  \label{f.spectra}
\end{figure}

A question arises as to whether such differences in the flux, photosphere, and surface
density profiles affect the resulting disk spectrum. In Fig.~\ref{f.spectra}
we present spectral profiles and their ratios 
(2-D to polytropic) for two accretion rates.
The spectra were calculated with ray-tracing routines \citep{bursa.raytracing} 
 using the BHSPEC package \citep{davisomer05}, assuming the inclination angle $i=70^o$ 
and distance to the observer $d=10\,{\rm kpc}$.
As BHSPEC gives tabulated solutions of the full,
frequency-dependent, radiative transfer
equations for the disk vertical structure taking the Compton scatterings into account
 in the disk atmosphere, the spectra
presented in  Fig.~\ref{f.spectra}
are not those of a simple multi-color blackbody. However, one should
be aware that using BHSPEC for calculating spectral color 
correction is not consistent with the assumed vertical structure
(calculated or height-averaged) because it is based on a stand-alone
disk model.

The general shape of the spectra is similar for both types of slim
disk models, because the emission profiles nearly coincide. However, the
spectra are not identical. 
The bottom panel of Fig.~\ref{f.spectra} presents ratios
of the spectral profiles of the corresponding solutions for each accretion rate.
They all coincide att low energies ($<0.1{\rm keV}$), while for 
higher energies the discrepancies
are as large as $3\%$ for  $1.0 \mdot_{\rm Edd}$ at $5{\rm keV}$. These
differences are atributed to (slightly) different profiles of the flux, 
the photosphere, and the surface density. We conclude that the proper
treatment of the vertical structure hardly affects the spectra of 
slim disks for this range of accretion rates ($\dot M<\dot M_{Edd}$) and 
the viscosity parameter ($\alpha\leq0.01$).


We end this section by giving fitting formulae for $N$, $f_H$, and $f_F$,
which approximate the full numerical 2-D solutions described in this paper 
with a polytropic slim disk model. An advantage of using these formulae
lies in avoiding the need to perform time-consuming calculations
of the vertical structure and avoiding numerical problems connected to
interpolation in the vertical solutions grid. The formulae
for the polytropic model parameters for $\alpha=0.01$ are
\begin{eqnarray}\nonumber
\label{poli.fit}
 N&=&3.25\times{\cal S}_N\\ 
 f_H&=&0.63\times{\cal S}_H\\\nonumber
 f_F&=&0.94\times{\cal S}_F,
\end{eqnarray}
where the spin correction coefficients ${\cal S}_N$, ${\cal S}_H$, and ${\cal S}_F$ are given by
\begin{eqnarray}\nonumber\label{poli.fit2}
 {\cal S}_N&=&1+0.002~(6-r_{ms}/M),\\\nonumber
 {\cal S}_H&=&1+0.003~(6-r_{ms}/M),\\\nonumber
 {\cal S}_F&=&1+0.064~(6-r_{ms}/M).
\end{eqnarray}
Here, $r_{ms}$ is the radius of the marginally stable orbit.
For a nonrotating BH, at the radius $r=7M$
(corresponding to the highest disk effective 
temperature), the fitting formulae are accurate to 1\% for
the emitted flux, the photospheric height and the surface density. 

\section{Discussion}
\label{s.conclusions}

Motivated by a desire to explain and fit
the observed spectra of accreting black holes in binary systems to theoretical models,
we have developed a 2-D model of optically thick
slim disks. These should be particularly relevant to transient
binaries. 
In quiescence, their inner disk regions are described well by optically thin
advection-dominated accretion flows \citep[ADAFs; see e.g.,][]{lasotaetal-96,dubusetal-01}.
However, a few black hole systems (e.g., GRS 1915+105 and LMC X-3)
have been observed in 
thermal states corresponding to disk luminosities higher than 
$0.3 L_{\rm Edd}$ \citep{mcclintockremillard03, steineretal-10},
and modeling these require optically thick models going beyond the standard
thin disks.

In this work we present
a 2-D slim disk model, in which the radial and vertical structures
are coupled. 
Such an approach eliminates arbitrary factors that influence
solutions of the usual polytropic slim 
disk model. 
The results were obtained under two key assumptions:
an alpha disk was assumed (dissipation proportional to pressure),
with a uniform value of $\alpha$,
and the fraction of the generated entropy that is advected
was computed at every radius under the assumption that
this fraction does not vary with the height above the disk
plane (Eq.~\ref{vs.dFdz}).
Both of these assumptions seem arbitrary, and we can offer no physical
motivation for the (conventional) choice we made.

Under these assumptions and for the value $\alpha=0.01$ of the viscosity
parameter, we computed and presented the detailed
structure of 2-D slim disks, parametrized by the
mass accretion rate, and the two Kerr metric parameters, $M$ and $a$.
Somewhat surprisingly, the spectra observed at infinity from such disks
differ by only a few percent from those obtained from previously considered
slim disk models (in which the equations and structure
correspond to a height average over a polytropic atmosphere).
Such differences are unlikely to introduce any large corrections
to spin measurements based on X-ray continuum fits 
made with corresponding height-averaged
polytropic models of slim disks. However, already the latter produce 
significantly softer spectra  in the sub-Eddington regime
than the Novikov \& Thorne model. For high luminosities, fits
based on slim disk models may therefore provide higher values of the
black hole spin parameter than corresponding fits based on the
\texttt{kerrbb} model \citep[e.g.,][]{shafee-06}. This issue will
be discussed in detail in a forthcoming paper \citep{bursa-slimbb}.

One has to be aware that the model of vertical
structure presented here is only an approximation
of the real physical processes taking place in disk interiors.
The diffusion approximation and the convection
treatment in the mixing length approach are known to successfully
describe media with large effective optical depths but break down when
the disk becomes optically thin. We have shown that the effective optical depth of slim accretion
disks may drop below unity for super-Eddington luminosities 
and sufficiently high values of $\alpha$. For such conditions,
a more sophisticated model of radiation transfer should be 
implemented. However, for $\alpha\leq0.01$ and $L\leq L_{\rm Edd}$
the assumptions of this work are self-consistent.
 For higher values of $\alpha$ their range of
applicability is limited to lower luminosities
(e.g., to $0.5 L_{\rm Edd}$ for $\alpha=0.1$).
Kerr-metric slim disks with low effective optical depth
were discussed by \cite{beloborodov-98}, who finds them to be significantly
hotter than the optically thick ones.
We have already started implementing a radiative transfer scheme valid
for disks with small effective optical depths into the scheme introduced in this
work. It will be presented and discussed in a future paper.

Another remark is connected to the fact that one can expect winds to be blown out
of the disk surface at super-Eddington luminosities. Such a phenomenon
may significantly change the disk structure, e.g., its thickness. This feature of slim disks, not described in
our calculations, has recently been cleverly modeled by  \cite{nir}.

\acknowledgements{
This work was supported in part by Polish Ministry of Science grants N203
0093/1466, N203 304035, N203 380336, N N203 381436. JPL
acknowledges support from the French Space Agency CNES, MB from ESA
PECS project No. 98040. We thank the anonymous referee for valuable comments.
}

\bibliographystyle{apj}
\bibliography{15256}

\end{document}